# Hidden-Markov Program Algebra with iteration

Annabelle McIver,[*] Larissa Meinicke,[†] Carroll Morgan[‡]

Jan 2011

**Abstract**

We use Hidden Markov Models to motivate a quantitative compositional semantics for noninterference-based security with iteration, including a refinement- or "implements" relation that compares two programs with respect to their information leakage; and we propose a program algebra for source-level reasoning about such programs, in particular as a means of establishing that an "implementation" program leaks no more than its "specification" program.

This joins two themes: we extend our earlier work, having iteration but only *qualitative* [37], by making it quantitative; and we extend our earlier *quantitative* work [27] by including iteration.

We advocate stepwise refinement and source-level program algebra — both as conceptual reasoning tools and as targets for automated assistance. A selection of algebraic laws is given to support this view in the case of quantitative noninterference; and it is demonstrated on a simple iterated password-guessing attack.

## 1 Introduction: extant theory and practices

*Hidden Markov Models*, or *HMM*'s, extend Markov Processes by supposing that the process state is not directly visible: only certain observations of it can be made [22]. How *HMM*'s motivate a quantitative noninterference-security program semantics is our principal topic: the hidden state of the *HMM* has "high security" and the observations that the *HMM* allows are "low security."

*Program algebra* is the manipulation of program texts themselves, i.e. as syntax and according to algebraic rules laid down beforehand, with the aim of showing equivalence or ordering with respect to a so-called "refinement" relation (§2) between one program and another. That requires a semantics, and proofs of the elementary rules wrt. that semantics. Furthermore these rules must be preserved by context in order for true algebra to be possible: in programming semantics, that last is called *compositionality*. This represents an "up front" cost for reasoning about program behaviours. When that cost has been paid however, just once, then the benefits accrue forever after — every time an equality or refinement can be shown syntactically without "descending" into the semantics.

---
[*]Macquarie University, Australia: `annabelle.mciver@mq.edu.au`
[†]Maquarie University, Australia: `larissa.meinicke@mq.edu.au`
[‡]University of NSW, Australia: `carrollm@cse.unsw.edu.au`

The significance of *iteration* is that its proper treatment, via suprema of chains, makes interesting demands on the semantic machinery already set-up for straight-line, quantitative noninterference programs [2, 27].

Our first specific contribution extends an existing (but recent [27]) compositional semantics for straight-line quantitative noninterference security, one with a novel two-level "hyperdistribution" semantics, by showing how hypers (for short) –previously introduced without detailed motivation– are in fact directly suggested by the mathematical machinery of *HMM*'s (§3). Our second contribution adds iterating programs to that (§6), requiring thus a treatment of nontermination and fixed-points: this would be straightforward were it not for the fact that supremum-completeness, on which fixed-points' existence usually relies, does *not* appear to hold.

Our third contribution (§7) is to show how, in spite of the incompleteness, we can via a more-specialised "termination order" retain discrete distributions for the treatment of loops: that gives a simpler theory than (the more general) measures would require. Nevertheless, our further goal of extending compositional-closure [27] to iteration does seem to require measures: at that point, there is no escape (§12).

Our final contribution is a selection of algebraic laws (§9), and the treatment of an example (§10) illustrating the style of reasoning we hope they will facilitate.

## 2 Program algebra and refinement

*Algebra* is powerful, and it is general; and it is especially useful in program verification where algebra's feature of *compositionality* allows the reuse that simplifies verification tasks. *Program* algebra in particular provides equalities or refinements (see below) that, although proved in isolation between program fragments, can then be reused freely within arbitrary contexts, drastically simplifying correct-by-construction and/or post-hoc verification arguments.

A *refinement* ordering between programs is weaker than equality: it defines the relationship that must hold between specifications and their implementations in a given application domain [49, 33, 5]. In special applications, such as noninterference security, the refinement relation is adjusted –usually made more restrictive– to take further aspects into account: here it will be the possible release of high-level information. Thus secure refinement checks not only (non)termination, but also compares programs to see which one releases more information about hidden, high-security variables: it is more distinguishing than standard program refinement.

For example, take integer variables v,h and suppose that v is visible (low-security) whereas h is hidden (high-security). Furthermore, assume an attacker with "perfect recall," i.e. one who remembers visible variables' values even if they are subsequently overwritten. (We explain this assumption in §4.2 and §12 below.) Then we would expect the refinement (v:=h÷2; v:=v÷2) ⊑ v:=h÷4 but, crucially, not the reverse. On the left, observing the first assignment to v as well as the second (perfect recall) allows us to distinguish h=1 from h=3; but on the right we cannot do that. The right-hand program is a refinement of the left-hand one because it is more secure; with an appropriate security-refinement algebra we would show this syntactically (§9.3).

## 3 Hidden Markov models and hyper-distributions

### 3.1 Basic structure of *HMM*'s

A Hidden Markov Model comprises a set $\mathcal{X}$ of states, a set $\mathcal{Y}$ of observations, and two stochastic matrices $\boldsymbol{T}, \boldsymbol{E}$ [22]: the *transition* probabilities $\boldsymbol{T}$ give for any two states $x_{\{0,1\}} \in \mathcal{X}$ the (conditional) probability $\boldsymbol{T}(x_1|x_0)$ that a transition will end in final state $x_1$ given that it began in initial state $x_0$; and the *emission* probabilities give for any state $x_0$ and observation $y_1 \in \mathcal{Y}$ the probability $\boldsymbol{E}(y_1|x_0)$ that $y_1$ will be emitted, and thus observed, given the initial state $x_0$. Typically an *HMM* is analysed over a number of steps $i = 0, 1, \cdots$ from some initial distribution $X_0$ over $\mathcal{X}$, so that a succession of states $x_1, x_2, \cdots$ and observations $y_1, y_2, \cdots$ occurs, where each $x_i$ related to $x_{i+1}$ by $\boldsymbol{T}$ and to $y_{i+1}$ by $\boldsymbol{E}$.

We assume finitely many states in the state space, and thus use discrete distributions throughout.[1]

We illustrate a single step in Fig. 1. With $X_0$ the distribution of incoming state $x_0$, the distribution $X_1$ of outgoing states $x_1$ is the multiplication of $X_0$ as a row-vector by $\boldsymbol{T}$ as a matrix, that is $\Pr(X_1{=}x_1) := \sum_{x_0} \Pr(X_0{=}x_0) \boldsymbol{T}(x_1|x_0)$. Similarly the distribution $Y_1$ of observations $y_1$ is given by a (matrix) multiplication amounting to $\Pr(Y_1{=}y_1) := \sum_{x_0} \Pr(X_0{=}x_0) \boldsymbol{E}(y_1|x_0)$. The "hidden" essence of the model is that though we cannot see the incoming $x_0$'s and the outgoing $x_1$'s directly, still the observation of $y_1$ tells us something about each if we do know the incoming state distribution $X_0$ and the matrices $\boldsymbol{T}, \boldsymbol{E}$.

### 3.2 *A priori* and *a posteriori* distributions on the state-space $\mathcal{X}$

The *a priori* distribution on the *HMM*'s input is $X_0$, and the *a posteriori* distribution on the input can be calculated from $\boldsymbol{T}, \boldsymbol{E}$, and $X_0$ in the usual way, via Bayes' formula, once we have observed $y_1$.

But we concentrate instead on the output. The *a priori* distribution of outgoing $x_1$ is $X_1$ as calculated in §3.1 above. Its *a posteriori* distribution is

---

[1] The countably infinite supports mentioned in the introduction occur, eventually, in spite of this finiteness assumption (§6.1).

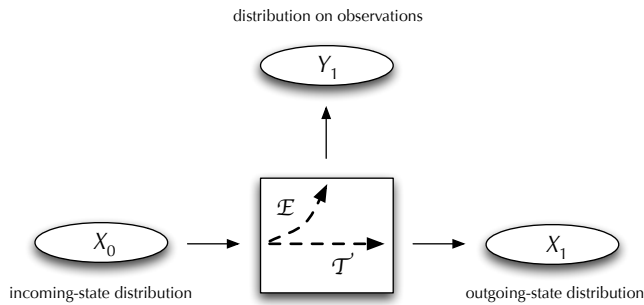

distribution on observations

incoming-state distribution    outgoing-state distribution

By *a priori* we mean that the distribution $X_1$ is determined statically, from information "already" available and in particular is not derived from an actual execution.

Figure 1: A Hidden Markov model, *a priori* view

conditioned on the emitted $y_1$ actually observed: it too is determined by the usual Bayes formula

$$\Pr(X_1{=}x_1|Y_1{=}y_1) \quad := \quad \frac{\sum_{x_0} \Pr(X_0{=}x_0)\boldsymbol{E}(y_1|x_0)\boldsymbol{T}(x_1|x_0)}{\sum_{x_0} \Pr(X_0{=}x_0)\boldsymbol{E}(y_1|x_0)} \ , \qquad (1)$$

that is the (joint) probability that $x_1, y_1$ both occurred divided by the overall (marginal) probability that $y_1$ occurred. Thus *before* we observe any $y_1$ we believe the distribution of outgoing $x_1$ to be $X_1$, and *after* we observe $y_1$ we believe that distribution to be as (1). This view is illustrated in Fig. 2.

### 3.3 The attacker's point of view: an equivalent representation

Although the matrices $\boldsymbol{T}, \boldsymbol{E}$ determine the *HMM* completely, we suggest that from the point of view of an attacker trying to determine the state of the *HMM*, it would be more useful to consider a different (but equivalent) formulation: the effect of one step from a known initial distribution $X_0$ is a joint distribution over observations in $\mathcal{Y}$ and their corresponding outgoing conditional distributions over $\mathcal{X}$: this structure thus comprises values $\Delta$ of type $\mathbb{D}(\mathcal{Y}{\times}\mathbb{D}\mathcal{X})$, where we write $\mathbb{D}\mathcal{X}$ and similar for the type of *discrete distributions* over $\mathcal{X}$, thus one-summing functions of type $\mathcal{X}{\to}[0,1]$. That is, each $\Delta$ gives for a pair $(y_1, \delta_1)$ in $\mathcal{Y}{\times}\mathbb{D}\mathcal{X}$ the probability that an attacker will observe $y_1$ and will conclude from it that $x_1$ has *a posteriori* distribution $\delta_1$.

We call such $\Delta$-values *hyperdistributions*, or just *hypers*. Since $\Delta$ is a joint distribution (jointly over $\mathcal{Y}$ and $\mathbb{D}\mathcal{X}$), we can speak of its left- and right-marginal distributions: the left-marginal distribution $\overleftarrow{\Delta}$ is of type $\mathbb{D}\mathcal{Y}$, and is in fact just $Y_1$ from above. That is, the distribution $Y_1$ of emitted observations is recovered as $\overleftarrow{\Delta}$.

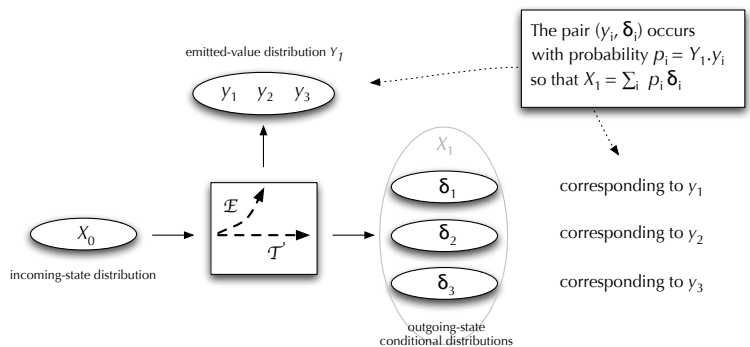

By *a posteriori* we mean that the conditional distributions $\delta_{\{1,2,3\}}$ are deduced *after* observation of the emitted values $y_{\{1,2,3\}}$, and represent a revision of the *a priori* knowledge of the outgoing state as represented in the $X_1$ of Fig. 1.

Figure 2: A Hidden Markov Model, *a posteriori* view

The right-marginal distribution $\vec{\Delta}$ of the hyper is more interesting: it is of type $\mathbb{D}^2\mathcal{X}$ and, although it *averages* to the outgoing state distribution $X_1$ (in the sense shown in Fig. 2), most of the popular (conditional) information-entropy measurements are likely to decrease, becoming less than the entropy of $X_1$ itself: that decrease quantifyies the "leak" that the emissions of $Y_1$ represent. For example, the *conditional Shannon Entropy* of $\vec{\Delta}$, defined $\sum_{\delta:\,\mathbb{D}\mathcal{X}} \vec{\Delta}(\delta)\mathrm{H}(\delta)$ over the possible *a posteriori* distributions $\delta$ is no more than $\mathrm{H}(X_1)$, the Shannon Entropy of the *a priori* outgoing distribution $X_1$ itself.[2]

Thus the denotational-style semantic representation we extract from *HMM*-theory is the *hyperdistribution* of type $\mathbb{D}(\mathcal{Y}\times\mathbb{D}\mathcal{X})$, a nesting of one distribution within another. As we will see, this allows us to equip the semantic space with a "refinement" partial order; but it is *security* refinement, so that for hypers $\Delta_{\{0,1\}}$ one can speak of whether $\Delta_0$ is more- or less secure than $\Delta_1$ or, if not, whether they are perhaps simply security-incomparable.

### 3.4 A probabilistic monad

A further benefit of conventional denotational techniques is our access to computational monads [15, 32, 47], simplifying the presentation considerably.

From here on, we use a dot "." for function application, rather than parentheses $(\cdot)$, writing thus $f.x$ rather than $f(x)$. For Curried functions we will usually have $f.x.y$ rather than e.g. either $f(x,y)$ or $f(x)(y)$.[3] As a result, given distri-

---

[2]For distribution $X$ in $\mathbb{D}\mathcal{X}$ such that $\Pr(X{=}x_i)$ is $p_i$, the Shannon Entropy $\mathrm{H}(X)$ of $X$ is given by $-\Sigma_i\, p_i \ln p_i$. As remarked, a number of other security-based definitions of entropy give the same inequality [24].

[3]An advantage of this is that it distinguishes function application from the many other uses of parentheses, and produces self-contained expressions thus of less clutter. In this respect we compare $\mathrm{H}.X = -\Sigma_x X.x \ln(X.x)$ with the conventional presentation of Shannon Entropy in

bution $X\colon\mathbb{D}\mathcal{X}$ the probability it assigns to $x\colon\mathcal{X}$ is simply $X.x$, that is $\Pr(X{=}x)$ but written more compactly and taking advantage of the fact that $X$ is just a function (of type $\mathcal{X}{\to}[0,1]$). The same economy accrues for random variables.

The monad structure for computations [32] supposes a triple $(\mathbb{K},\boldsymbol{\eta},\boldsymbol{\mu})$ where $\mathbb{K}$ is an endofunctor in a given category, and $\boldsymbol{\eta},\boldsymbol{\mu}$ are natural transformations satisfying certain coherence conditions. An example of this is the *Giry monad* [15], typically used for probabilistic computations; in its general form, its functor takes an object $(\Omega,\mathcal{B}_\Omega)$ comprising a set $\Omega$ and a sigma-algebra $\mathcal{B}_\Omega$ on it to the set of probability measures on $(\Omega,\mathcal{B}_\Omega)$, endowed with a suitable sigma-algebra of its own, induced from the given $\mathcal{B}_\Omega$.

Working here with discrete measures, our use of the monad will be modest and we will use suggestive names for its components, based on its specialisation to discrete distributions and functional programming. In particular,

**functor $\mathbb{D}$** – Given set $SS$ write $\mathbb{D}SS$ for the set of discrete distributions over $SS$.

**push-forward map** – Given two sets $\mathcal{X},\mathcal{Y}$ and a function $f\colon\mathcal{X}{\to}\mathcal{Y}$ write $\mathbb{D}f$, the action of the functor on the function, as $\mathsf{map}.f\colon\mathbb{D}\mathcal{X}{\to}\mathbb{D}\mathcal{Y}$.[4] In the probability literature this is called the *push-forward*, defined for any $X\colon\mathbb{D}\mathcal{X}$ and $y\colon\mathcal{Y}$ in the discrete case as

$$\mathsf{map}.f.X.y \quad := \quad X.(f^{-1}.y) \ = \ (\ \textstyle\sum x\colon\mathcal{X}\mid f.x{=}y\bullet X.x\ )\ .$$

**multiplication avg** – The multiplication (natural) transformation $\boldsymbol{\mu}\colon\mathbb{D}^2\mathcal{X}{\to}\mathbb{D}\mathcal{X}$ averages the distributions in its argument distribution-of-distributions, to give a distribution again. We write that as $\mathsf{avg}$ for "average" and in the discrete case for $\boldsymbol{X}\colon\mathbb{D}^2\mathcal{X}$ it is defined for any $x\colon\mathcal{X}$ as

$$\mathsf{avg}.\boldsymbol{X}.x \quad := \quad (\ \textstyle\sum X\colon\mathbb{D}\mathcal{X}\bullet \boldsymbol{X}.X{\times}X.x\ )\ .$$

**Kleisli composition via lifting** For two functions $f\colon\mathcal{X}{\to}\mathbb{D}\mathcal{Y}$ and $g\colon\mathcal{Y}{\to}\mathbb{D}\mathcal{Z}$, the *lift* of $g$, written $g^*$, is defined to be the functional composition $\mathsf{avg}\circ\mathsf{map}.g$ of type $\mathbb{D}\mathcal{Y}{\to}\mathbb{D}\mathcal{Z}$ so that the Kleisli composition $g$ after $f$ is expressed $g^*{\circ}f$ in the usual way.

Similar definitions and notations apply for the monad $\underline{\mathbb{D}}$ associated with the *partial distributions* that sum to no more than one [21, 47].

The immediate benefit from this monadic structure is the sequential composition of two *HMM*'s, written say $H_1;H_2$, so that the final *a posteriori* distribution takes the observations from the first as well as the second *HMM* observables into account. If we take the type of $H_{\{1,2\}}$ to be $\mathcal{Y}{\times}\mathbb{D}\mathcal{X} \to \mathbb{D}(\mathcal{Y}{\times}\mathbb{D}\mathcal{X})$, then we want our definition of $H_1;H_2$ to have similar type, thus giving it the same features as a single *HMM* provided that the observations from both of its components are considered: the general case is illustrated in Fig. 3.

---

Footnote 2 with its indices $i$ and temporary names $p_i$.

[4]This is consistent with the definition of $\mathsf{map}$ in functional programming.

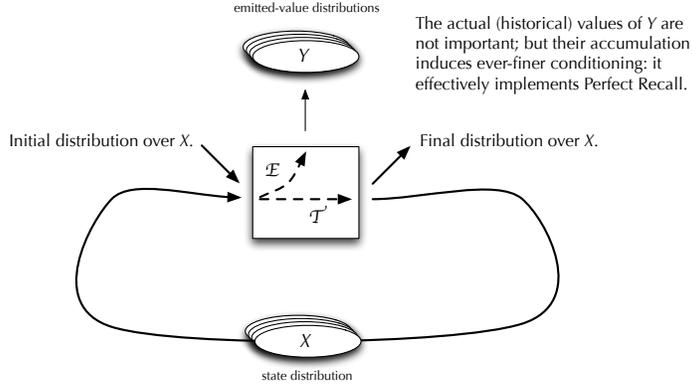

Figure 3: A Hidden Markov Model: iteration accumulates leakage

For our single composition $H_1; H_2$, the outgoing result of $H_1$ is an *a posteriori* hyper which is then presented "en bloc" as input to $H_2$. Via the type constructor, that intermediate hyper is a partitioning of some flattened distribution of type $\mathcal{Y} \times \mathbb{D}\mathcal{X}$ according to the observables emitted by $H_1$;[5] each partition of that flattened distribution –itself of type $\mathcal{Y} \times \mathbb{D}\mathcal{X}$– is separately input to $H_2$, but after the final output is produced the partitioning is "reassembled" in the overall the final *a posteriori* distribution, thus neatly taking both the observations from $H_{\{1,2\}}$ into account. Crucially this partitioning and reassembling is done according to the original weightings, which is what allows us to use the monad:

$$\mathcal{Y} \times \mathbb{D}\mathcal{X} \xrightarrow{H_1} \mathbb{D}(\mathcal{Y} \times \mathbb{D}\mathcal{X}) \xrightarrow{\mathsf{map}.H_2} \mathbb{D}^2(\mathcal{Y} \times \mathbb{D}\mathcal{X}) \xrightarrow{\mathsf{avg}} \mathbb{D}(\mathcal{Y} \times \mathbb{D}\mathcal{X}) \ .$$

Here the original input type $\mathcal{Y} \times \mathbb{D}\mathcal{X}$ is transformed as we suggest by $H_1$ to $\mathbb{D}(\mathcal{Y} \times \mathbb{D}\mathcal{X})$, which then in its partitioned form is passed to $H_2$ and, via the map/avg construction, that partition is reassembled after the action of $H_2$ on its components. That supplies our definition for $H_1; H_2$, thus also of type $\mathcal{Y} \times \mathbb{D}\mathcal{X} \to \mathbb{D}(\mathcal{Y} \times \mathbb{D}\mathcal{X})$. Note we do not need $\mathcal{Y}^2$ to "combine" the observations of the two separate *HMM*'s, an important advantage of this presentation: in §4.2(2,4) this is explained further.

---

[5] We call it a distribution simply to avoid a proliferation of names. In fact it is isomorphically a distribution of type $\mathbb{D}(\mathcal{Y} \times \mathcal{X})$ whose left marginal is a point distribution.

# 4 Quantitative noninterference security for programs

## 4.1 Noninterference via hidden- and visible variables; atomicity

Take a simple programming model comprising a finite set $\mathcal{H}$ of hidden states, ranged over by (high-security) program variable(s) h, and a finite set $\mathcal{V}$ of visible states, ranged over by (low-security) program variable(s) v. The state space overall is thus the product $\mathcal{V} \times \mathcal{H}$, and our program texts refer to variables v, h.[6]

Observers of the program's execution can see v, but they cannot see h. Attackers of the program try to learn about h's final values, or at least their distribution, by observing v's values as execution of the program proceeds.

We begin with assignment statements as a basis: a simultaneous assignment is written v, h:= $V, H$, allowing both expressions $V, H$ to refer to the initial values of v, h without worrying about which one is updated first: the two expressions $V, H$ may contain variables of either kind, or both. We base the assignment on a probabilistic-choice syntax x:$\in X$ that means "choose the new value of $x$ according to the distribution $X$," where $X$ itself is a distribution-valued expression that possibly depends on the initial value of x as well as other variables.

To keep track of variables in the generic semantic definitions, we write the distribution-valued expressions as explicit functions applied to them, so arriving at v, h:$\in$ $E$.v.h, $T$.v.h as our basic simultaneous probabilistic assignment to the two variables; actual program texts of course simply use expressions over v, h in which such functions might occur. Thus $E$.v.h is a distribution, depending on the initial values of v, h, according to which v's new value is chosen. Similarly $T$.v.h is the distribution for the choice of h's new value. The statement is *atomic* in the sense that only its results are accessible, not how they were computed.

Ordinary, non-probabilistic assignments can be written v, h:= $E$.v.h, $T$.v.h in which $E, T$ now give values rather than distributions; they are clearly the special case of the above for point distributions, and so do not need a separate treatment.

## 4.2 Connecting *HMM*'s and the programming model; perfect recall

The connection is made by identifying $\mathcal{V}$ with $\mathcal{Y}$, and $\mathcal{H}$ with $\mathcal{X}$. The visible v corresponds to the emitted observations $y$, thus to a sort of "output buffer"; and the hidden h corresponds to the state $x$, passed from one program fragment through sequential composition to the next one.

---

[6]For simplicity we are assuming that multiple hidden- or visible variables are collected separately within vectors h or v, i.e. that v, h are the only variables present; but we won't clutter the presentation with the "overhooks" for that. The program texts can refer of course to individual elements of the vectors, which references are interpreted as projections etc. in the usual way.

A slight generalisation is that the probabilistic choices can be influenced by the immediately previous emitted value (by $y$ from the previous step, whose value is the initial value of v for this step), whereas in an *HMM* this is typically not done. This is only a notational convenience, since clearly the *HMM*'s state spaces can be elaborated to allow the same freedom; but such conveniences are a part of adapting the *HMM* framework to programming practice.

Further elaborating the adaptation, we make the following remarks:

1. A typical sequential program will execute many individual atomic steps successively. The outgoing state from one step will be both its final value of h, fed-in automatically as the incoming state of the next step, *and* the last emitted observable value, found in v.

2. Although the observations emitted from each step will successively overwrite earlier values in v, the conditioning observations of those earlier outputs caused is *not* lost: it is preserved by the map/avg composition. Thus the partitioning expressed by the growing support-set of the outer $\mathbb{D}$ becomes finer on each step, so that deductions made by an attacker's having seen an earlier v are never forgotten. This is called *Perfect Recall* [18].

3. The distribution $E$.v.h from which v's final value is chosen corresponds to the stochastic matrix $\boldsymbol{E}$ of the *HMM*. In effect, the h in $E$.v.h selects the row of $\boldsymbol{E}$ that gives the distribution from which $y$, that is from which v is chosen. Similarly, the distribution $T$.v.h from which h's final value is chosen corresponds to the stochastic matrix $\boldsymbol{T}$. The programs' access to v is why we include $(\mathcal{Y}\times)$ in the state.

4. The *a priori* view of the program is the extent to which we can determine the distribution of the final values of h by knowing the incoming distribution of v, h and the program text. The *a posteriori* view reflects the extra information about h finally that we have once we actually execute the program and note the successive emissions in v that occur during that execution. However the *values* of those emissions need not be remembered: only the conditioning they induce is important. That is why we do not need "sequence of $\mathcal{Y}$" in our state, in spite of perfect recall: the recall is expressed in the outer $\mathbb{D}$.

## 5 *Refinement* increases entropy compositionally

Our advocacy of stepwise refinement [49] for development of quantitatively non-interference secure programs suggests comparisons of specifications $S$ with implementations $I$. Say that $S$ is "Shannon-refined" by $I$, writing $S \preceq_{\text{se}} I$, just when for every incoming distribution of hidden values h the *a posteriori* conditional Shannon Entropy produced by $I$, for h, is at least that produced by $S$ (as at 4. above). Stepwise refinement wrt Shannon Entropy requires transitivity of ($\preceq_{se}$) obviously; but it also would require that $S \preceq_{se} I$ imply $\mathcal{C}(S) \preceq_{se} \mathcal{C}(I)$ for any context $\mathcal{C}$ — that is, it should be *compositional*. And it is not, in general.

Define analogously ($\preceq_{br}$) for comparing conditional Bayes Risk of outputs, in the same way; it is not compositional either.[7]

The refinement relation ($\sqsubseteq$) introduced earlier [27], and extended here for iteration, in fact *is* compositional; and furthermore, it implies both ($\preceq_{se}$) and ($\preceq_{br}$). (Counter-examples for compositionality of ($\preceq_{se}$) and ($\preceq_{br}$) are given in the in the extended version of that work.) We now explain refinement.

## 5.1 Comparing hyperdistributions

We begin for simplicity with an entirely hidden state $\mathcal{X}$ (i.e. without $\mathcal{Y}$) thus having hypers $\mathbb{D}^2\mathcal{X}$. We ask whether, for (each) fixed observation $y_1$, one *HMM* "reveals more" than the other in a sense made precise as follows.

A hyper $\Delta_S$ in $\mathbb{D}^2\mathcal{X}$, produced say as the output of one *HMM*, is "refined by" another hyper $\Delta_I$, produced by another *HMM* of the same type, if two distributions $\delta_S^{\{1,2\}}:\mathbb{D}\mathcal{X}$ in the support of $\Delta_S$ can be merged to form a single distribution $\delta_I$ in what becomes $\Delta_I$. This merging increases a variety of (conditional) entropies, including the two mentioned above. We say that one *HMM* is entropy-refined by another when for corresponding inputs and corresponding values of emitted observables their outgoing hypers are refinement-related.

For an example, we restrict to Booleans $\mathsf{T}, \mathsf{F}$, and write $\{\!\{x^{@p}, y^{@q}, \cdots, z^{@r}\}\!\}$ for the discrete distribution assigning probabilities $p, q, \cdots, r$ to values $x, y, \cdots, z$: it is partial or total depending on whether $p+q+\cdots+r$ equals 1. Suppose the specification hyper $\Delta_S$ contains two distributions $\delta_S^1 := \{\!\{\mathsf{T}^{@\frac{1}{3}}, \mathsf{F}^{@\frac{2}{3}}\}\!\}$ and $\delta_S^2 := \{\!\{\mathsf{T}^{@\frac{1}{2}}, \mathsf{F}^{@\frac{1}{2}}\}\!\}$ with probabilities $p^1 := 1/4$ and $p^2 := 1/3$ respectively: thus $\Delta_S$ is partial and can itself be written $\{\!\{\delta_S^{1\ @\frac{1}{4}}, \delta_S^{2\ @\frac{1}{3}}, \cdots\}\!\}$. We first calculate a weighted merge as follows:

- Scale $\delta_S^{\{1,2\}}$ by their respective probabilities $p^{\{1,2\}}$ in $\Delta_S$ to get partial distributions $\{\!\{\mathsf{T}^{@\frac{1}{12}}, \mathsf{F}^{@\frac{1}{6}}\}\!\}$ and $\{\!\{\mathsf{T}^{@\frac{1}{6}}, \mathsf{F}^{@\frac{1}{6}}\}\!\}$.

- Add those together pointwise to get $\{\!\{\mathsf{T}^{@\frac{1}{4}}, \mathsf{F}^{@\frac{1}{3}}\}\!\}$.

- Normalise to get $\delta_I := \{\!\{\mathsf{T}^{@\frac{3}{7}}, \mathsf{F}^{@\frac{4}{7}}\}\!\}$ with probability $p := 7/12$ in $\Delta_I$.

Then we refine $\Delta_S$ by removing the two distributions $\delta_S^{\{1,2\}}$ (total weight 7/12) and replacing them by their weighted merge, the single $\delta_I$ (of the same weight), to give $\Delta_I$. All the other points in the support of $\Delta_S$ would carry over unchanged into $\Delta_I$; but of course this process can be repeated, since refinement is to be transitive. We see at (2) below that general entropy refinements are achieved by merges of more than two sources, having multiple targets and by "pre-splitting" sources proportionally to allow them to participate in more than one merge: the essential idea is as given here.

---

[7] The Bayes Risk is the largest guaranteed chance that one guess of $\mathsf{h}$ is *incorrect*.

## 5.2 Preliminary defintion of entropy refinement

Distributions over our states in $\mathcal{X}$, i.e. in $\mathbb{D}\mathcal{X}$, are called *inner* distributions or just "inners." Distributions over inners, i.e. in $\mathbb{D}\mathbb{D}\mathcal{X} = \mathbb{D}^2\mathcal{X}$, are hypers, as we have seen. If we want to concentrate on the "outer" $\mathbb{D}$ of a hyper, we refer to that as the *outer* distribution, or just "outer." We will (briefly) need distributions of hypers $\mathbb{D}^3\mathcal{X}$, called *super* distributions or just "supers."

**Definition 5.1** *Entropy refinement (preliminary definition)* Let the state-space be $\mathcal{X}$, a finite set, and consider two hypers $\Delta_{\{S,I\}}\colon\mathbb{D}^2\mathcal{X}$. We say that $\Delta_S$ is *entropy refined* by $\Delta_I$, written $\Delta_S \preceq \Delta_I$, iff there is a super $\boldsymbol{\Delta}\colon\mathbb{D}^3\mathcal{X}$ such that

$$\Delta_S = \mathsf{avg}.\boldsymbol{\Delta} \qquad \text{and} \qquad \mathsf{map.avg}.\boldsymbol{\Delta} = \Delta_I \ . \qquad \square$$

We return to our example, hyper $\Delta_S$ now with three inners $\delta_S^1 := \{\!\{\mathsf{T}^{@\frac{1}{3}}, \mathsf{F}^{@\frac{2}{3}}\}\!\}$ and $\delta_S^2 := \{\!\{\mathsf{T}^{@\frac{1}{2}}, \mathsf{F}^{@\frac{1}{2}}\}\!\}$ and $\delta_S^3 := \{\!\{\mathsf{T}^{@1}\}\!\}$ with probabilities $p^1 := 1/4$ and $p^2 := 1/3$ and $p^3 := 5/12$ respectively, where the third inner is chosen to bring the (outer's) sum to 1, i.e. to make it total.

Now to reach the entropy refinement of $\Delta_S$ given by hyper $\Delta_I$, we merge the first two inners and simply carry the third through. The mediating super $\boldsymbol{\Delta}$ contains the two hypers

- hyper $\Delta^1 := \{\!\{\delta_S^{1\ @\frac{3}{7}}, \delta_S^{2\ @\frac{4}{7}}\}\!\}$ with probability 7/12 in $\boldsymbol{\Delta}$, and

- hyper $\Delta^2 := \{\!\{\delta_S^{3\ @1}\}\!\}$ with probability 5/12 in $\boldsymbol{\Delta}$,

so that $\Delta_S = \mathsf{avg}.\boldsymbol{\Delta}$, for example because

$$\mathsf{avg}.\boldsymbol{\Delta}.\delta_S^1 \quad = \quad 3/7 \times 7/12 \quad = \quad 1/4 \quad = \quad p_1 \quad = \quad \Delta_S.\delta_S^1 \ .$$

From Def. 5.1 the hyper $\Delta_I$ is therefore given by $\mathsf{map.avg}.\boldsymbol{\Delta}$, that is

- inner distribution $\mathsf{avg}.\Delta^1 = \mathsf{avg}.\{\!\{\delta_S^{1\ @\frac{3}{7}}, \delta_S^{2\ @\frac{4}{7}}\}\!\} = \{\!\{\mathsf{T}^{@\frac{3}{7}}, \mathsf{F}^{@\frac{4}{7}}\}\!\}$
  with probability 7/12, and

- inner distribution $\mathsf{avg}.\Delta^2 = \mathsf{avg}.\{\!\{\delta_S^{3\ @1}\}\!\} = \delta_S^3$ itself, carried through
  with probability 5/12 as we expected.

A second example of entropy refinement is given at (2) below.

It can be shown [27] that refinement is indeed a partial order: reflexivity is obvious, anti-symmetry follows from an entropy-based argument or alternatively from "colour mixing" [45]. Its transitivity can be shown using matrices, or by a monadic approach using general properties of $\mathsf{map}$ and $\mathsf{avg}$ and specific properties of the probabilistic functor.

# 6 Iteration, refinement chains and incompleteness

Iteration and entropy refinement taken together impose new demands on our semantic space, closure under limits: iterations are usually defined via least fixedpoints whose existence is trivial if the program space forms a *cpo* under the refinement ordering [46]. Since we have not yet introduced non-termination, the space $(\mathbb{D}^2\mathcal{X}, \preceq)$ has no least element, and so it is not a *cpo*. But it is not a *dcpo* either, as we show below: not all of its non-empty chains have a supremum.

As a result, the usual technique of defining iterations via refinement-least fixedpoints will not obviously apply –even after extending the space and its ordering to incorporate non-terminating behaviours– and we will have to do something slightly different (§7.3).

## 6.1 An example of incompleteness

Define again $\mathcal{X}:=\{\mathsf{T},\mathsf{F}\}$, and let $\delta_p$ be the inner $\{\!\{\mathsf{T}^{@p}, \mathsf{F}^{@1-p}\}\!\}$, alternatively written $\mathsf{T}\ _p\!\oplus\mathsf{F}$, for any $0\le p\le 1$. Form the sequence of hypers

$$
\begin{aligned}
\Delta_1 &:= \{\!\{\delta_0^{@\frac{1}{2}}, \delta_1^{@\frac{1}{2}}\}\!\} \\
\Delta_2 &:= \{\!\{\delta_0^{@\frac{1}{4}}, \delta_{1/2}^{@\frac{1}{2}}, \delta_1^{@\frac{1}{4}}\}\!\} \\
\Delta_3 &:= \{\!\{\delta_0^{@\frac{1}{8}}, \delta_{1/4}^{@\frac{1}{4}}, \delta_{1/2}^{@\frac{1}{4}}, \delta_{3/4}^{@\frac{1}{4}}, \delta_1^{@\frac{1}{8}}\}\!\} ,\quad \cdots,
\end{aligned}
\tag{2}
$$

in $\mathbb{D}^2\mathcal{X}$ whose pattern should be evident.

From Def. 5.1 we see that each of these hypers is an entropy refinement of the preceding: for example to get from $\Delta_2$ to $\Delta_3$ we first "pre-split" $\Delta_2$ into smaller pieces

$$
\{\!\{\delta_0^{@\frac{1}{8}}, \underbrace{\delta_0^{@\frac{1}{8}},\quad \delta_{1/2}^{@\frac{1}{8}}}_{\text{merge}}, \delta_{1/2}^{@\frac{1}{4}}, \underbrace{\delta_{1/2}^{@\frac{1}{8}},\quad \delta_1^{@\frac{1}{8}}}_{\text{merge}}, \delta_1^{@\frac{1}{8}}\}\!\}
$$
$\downarrow \qquad\qquad\qquad \downarrow \qquad\qquad\qquad \downarrow$

and then merge the selected inners as explained above, that is

$$
\delta_0^{@\frac{1}{8}} + \delta_{1/2}^{@\frac{1}{8}} \;=\; \delta_{1/4}^{@\frac{1}{4}} \quad\text{and}\quad \delta_{1/2}^{@\frac{1}{8}} + \delta_1^{@\frac{1}{8}} \;=\; \delta_{3/4}^{@\frac{1}{4}}
$$

to give $\Delta_3$ when we allow the un-merged distributions simply to carry through.[8]

Now by symmetry any hyper $\Delta$ that was a refinement-limit of the chain (2) would have to be uniform (except for the endpoints) but with a countably infinite support, since the supports of the chains' *elements* grow without bound — and uniform, infinite and discrete distributions do not exist. The actual limit of that refinement chain is in fact the *measure* over the distributions $\delta_p$ given by taking $p$ uniformly from $[0, 1]$, and that is outside our space $\mathbb{D}^2\mathcal{X}$. Writing $\mathbb{M}$ for "measure" (informally, i.e. without being specific about the sigma-algebra) we find our limit in $\mathbb{M}\mathbb{D}\mathcal{X}$ rather than $\mathbb{D}^2\mathcal{X}$.

## 6.2 Dealing with the incompleteness: proper measures

Pursuing the $\mathbb{MDX}$ strategy suggested above would lead us through steps like these:

1. Define a metric over $\mathbb{DX}$, i.e. provide a distance function between (discrete) distributions. For reasons we explain in §12 we would choose the Kantorovich metric [13] which is advocated for this kind of application anyway [47].

2. Generate the Borel algebra from the Kantorovich metric.

3. Define refinement between hypers that are proper measures, a generalisation of the "split/merge" of Def. 5.1 and explained in our previous work [27] for the discrete case.

4. Observe that the resulting, more general semantic space $\mathbb{DX} \to \mathbb{MDX}$ allows (still) a monadic treatment of sequential composition.

5. Define the program semantics §8 in that more sophisticated space.

But we do not do that here. Instead, in this report we limit our extensions to *just what will suffice* for the quantitative security of iterative programs, including making refinement-based comparisons between them, as part of our general programme of expanding the scope of this approach to deal with realistic situations. In fact, we will see that refinement chains generated by the fixed-point definition of loops *do* have a refinement-sup in our space — that is, "loop-approximant chains" are a strict subset of all possible refinement chains, and do not in particular contain examples like (2) above.

Thus we can remain within the space of discrete hypers $\mathbb{D}^2\mathcal{H}$, which allows a drastic simplification (compared with $\mathbb{MDH}$) of the presentation. How this is done is the topic of the next section: we will be using *partial* discrete distributions to represent nontermination, thus concentrating on a space $\mathcal{V} \times \mathbb{DH} \to \underline{\mathbb{D}}(\mathcal{V} \times \mathbb{DH})$ for denotations of programs.

---

[8]Using our formal definition Def. 5.1 introduces a super $\pmb{\Delta}$ to mediate the entropy refinement $\Delta_2 \preceq \Delta_3$; it is given by

$$\Delta_2 \left\{ \begin{array}{c} \left| \begin{array}{c} \delta_0^{@\frac{1}{4}} \\ \delta_{1/2}^{@\frac{1}{2}} \\ \delta_1^{@\frac{1}{4}} \end{array} \right| \end{array} \right. \xleftarrow{\text{avg}} \overset{\text{normalise the columns}}{\underset{}{\text{to give hypers of } \pmb{\Delta}}} \left| \begin{array}{c} \delta_0^{@\frac{1}{8}} \end{array} \right| \left\| \begin{array}{c} \delta_0^{@\frac{1}{8}} \\ \delta_{1/2}^{@\frac{1}{8}} \end{array} \right\| \left\| \begin{array}{c} \delta_{1/2}^{@\frac{1}{4}} \end{array} \right\| \left\| \begin{array}{c} \delta_{1/2}^{@\frac{1}{8}} \\ \delta_1^{@\frac{1}{8}} \end{array} \right\| \left| \begin{array}{c} \delta_1^{@\frac{1}{8}} \end{array} \right|$$

$$\downarrow \quad \downarrow \quad \downarrow \quad \downarrow \quad \downarrow \qquad \text{map.avg}$$

$$\delta_0^{@\frac{1}{8}} \quad \underbrace{\delta_{1/4}^{@\frac{1}{4}} \quad \delta_{1/2}^{@\frac{1}{4}} \quad \delta_{3/4}^{@\frac{1}{4}}}_{\Delta_3} \quad \delta_1^{@\frac{1}{8}}$$

The two merges of inners referred to in the main text occur in columns 2,4; the columns 1,3,5 are the inners that carry through unchanged from $\Delta_2$ to $\Delta_3$.

# 7 Semantics for the *HMM*-interpretation of secure iterating programs

## 7.1 Denotations of programs

Here we give a precise construction of a semantic space, and the interpretation of a small programming language in it. The language includes probability, visible vs. hidden variables, and iteration.

For *noninterference* we imagine a finite underlying state space of two parts, named $\mathcal{V}$ and $\mathcal{H}$ where $\mathcal{V}$ is the "visible" part of the state and $\mathcal{H}$ is its "hidden" part. Because the $\mathcal{H}$ part is hidden our underlying state space will not be simply the Cartesian product of those two components, but rather the set $SS := \mathcal{V} \times \mathbb{D}\mathcal{H}$ comprising the product of the visible part $\mathcal{V}$ (as is) and the *distributions* $\mathbb{D}\mathcal{H}$ over the hidden part.

For *nontermination* we consider program outputs to be of type $\underline{\mathbb{D}}SS$, that is $\underline{\mathbb{D}}(\mathcal{V} \times \mathbb{D}\mathcal{H})$, the *partial* distributions over $SS$ — this represents a slight generalisation of the type suggested above for programs in that the partiality (the one-deficit) is used to describe the probability of the program's failing to terminate [21, 20, 39, 28]. As before, we call elements of $\underline{\mathbb{D}}SS$ hypers, referring if necessary to *partial* hypers when the distinction is important. Thus $SS \to \underline{\mathbb{D}}SS$, that is $\mathcal{V} \times \mathbb{D}\mathcal{H} \to \underline{\mathbb{D}}(\mathcal{V} \times \mathbb{D}\mathcal{H})$ is the type we propose for programs: from an initial state $(v, \delta)$ in $SS$ a program determines a partial distribution $SS$, i.e. a distribution whose supports have structure $(v', \delta')$, as its final output.

Recall from §3.4 that we write function application as $f.x$, with "." associating to the left. Operators without their operands are written between parentheses, as $(\preceq)$ for example. [9]

## 7.2 The Entropy Refinement Order between programs

As usual our orders on programs will be the pointwise orders on their results. Our first order is the *Entropy Refinement* set out at Def. 5.1, adapted to deal with partial hypers and to take the $\mathcal{V}$ portion of the state into account.

**Definition 7.1** *Entropy refinement (generalising Def. 5.1)* Let the state-space be $SS = \mathcal{V} \times \mathbb{D}\mathcal{H}$, with $\mathcal{V}, \mathcal{H}$ both finite, and define $Q \colon SS \to \mathbb{D}(\mathcal{V} \times \mathcal{H})$ with $Q.(v, \delta).(v', h')$ equal to $\delta.h'$ if $v = v'$, otherwise zero. [10]

---

[9] The latter (known as *sections* in functional programming) allows us easily to write expressions relating operators themselves, such as the succinct $(<) \subseteq (\leq)$ stating that less-than is a subset of less-than-or-equals as a relation. Thus the former "dot" convention distinguishes function application from sections as well.

As a further example (though not needed in this report), as part of the definition of the Giry monad one defines *evaluation functions* $E_B$ that, given a measure $\mu$ as argument, return $\mu$ applied to the measurable set $B$ as the result. With sections and the "dot" convention one writes directly $(.B)$ for this function: the well established syntactical rules for sections then ensure that $E_B(\mu) = (.B).\mu = \mu.B$ automatically. A separate introduction, definition and explanation of the $E_B$ notation is not necessary.

[10] More succinctly, this is defining the product distribution $Q.(v, \delta) := \{\!\{v\}\!\} \times \delta$.

For two hypers $\Delta_{\{S,I\}}$: $\mathbb{D}SS$, we say that $\Delta_S$ is *entropy refined* by $\Delta_I$, writing $\Delta_S \preceq \Delta_I$, just when $\mathsf{map}.Q.\Delta_S \preceq \mathsf{map}.Q.\Delta_I$ according to our preliminary definition Def. 5.1 of entropy refinement, but taking our $\mathcal{V} \times \mathcal{H}$, here, all at once as just $\mathcal{X}$ there and generalising $\mathsf{map}, \mathsf{avg}$ to partial distributions in the obvious way. □

Like the preliminary definition, ($\preceq$) defines a partial order on hyper-distributions. Note that a consequence of this definition is that entropy refinement does not change the distribution of the visible variables: that is if $\Delta_S \preceq \Delta_I$, then in fact $\overleftarrow{\Delta}_S = \overleftarrow{\Delta}_I$ where we recall that $\overleftarrow{\Delta}$ is the left-marginal distribution of the product distribution $\Delta$: $\mathbb{D}(\mathcal{V} \times \mathbb{D}\mathcal{H})$. Similarly, the *a priori* distribution of $\mathsf{h}$ associated with each value of $\mathsf{v}$ is left unchanged.

We now address the incompleteness issue raised in §6.1.

### 7.3 The Termination Refinement Order between programs

We follow an approach that allows us to distinguish between chains produced by iteration and those produced by refinement more generally [41, 42]: we use a stronger order for which our space *is* complete.

For a partial hyper $\Delta$: $\mathbb{D}SS$, the probability that it terminates is just its total weight, written $\sum\Delta$; equivalently, the amount by which it fails to sum to 1 is its probability of nontermination. We define a partial order that allows increase of termination only, as follows:

**Definition 7.2** *Termination Refinement* For $\Delta_{\{S,I\}}$ in $\mathbb{D}SS$, we say that $\Delta_S$ is termination-refined by $\Delta_I$, written $\Delta_S \leq \Delta_I$, just when for all $s = (v, \delta)$ in $SS$ we have $\Delta_S.s \leq \Delta_I.s$. This is simply the pointwise extension of ($\leq$) on the real-valued probabilities. □

Our space is trivially closed under sup-chains in this termination order, since the probabilities themselves are bounded above (by 1). We will in due course show that the fixed-point definition of iteration generates termination chains, and so the completeness here will give us just the well definedness we need. That is, we will rely on

**Lemma 7.1** *Termination completeness* Let $\Delta_0 \leq \Delta_1 \leq \cdots$ be a ($\leq$)-chain of hypers in $\mathbb{D}SS$. Then the chain has a ($\leq$)-least upper bound $\bigvee_i \Delta_i$ in $\mathbb{D}SS$.
*Proof:* Completeness of $[0, 1]$. □

Note that everywhere-terminating programs are maximal in this *cpo*.

### 7.4 Secure refinement between programs

The primary order of interest on our space, secure refinement, allows both entropy refinement and termination refinement:

**Definition 7.3** *Secure Refinement*   Given (partial) hypers $\Delta_{\{S,I\}}$ in $\mathbb{D}SS$, we define *Secure Refinement* as the composition of the two other orders: first termination refinement, and then entropy refinement. We have $\Delta_S \sqsubseteq \Delta_I$ just when there is an intermediate hyper $\Delta$ such that $\Delta_S \leq \Delta$ and $\Delta \preceq \Delta_I$.   □

Observe trivially that ($\leq$) is a strengthening of ($\sqsubseteq$), by reflexivity of ($\preceq$). Like termination and entropy refinement, secure refinement it is a partial order on hyper-distributions. Reflexivity holds trivially from that of ($\leq$) and ($\preceq$). The transitivity of ($\sqsubseteq$) follows from transitivity of the two other orders, plus the fact that ($\sqsubseteq$) $\supseteq$ ($\preceq$)∘($\leq$). For antisymmetry we reason that if $A \sqsubseteq C$ and $C \sqsubseteq A$ then there must exist a $B$ and $D$ such that $A \leq A+B \preceq C$ and $C \leq C+D \preceq A$. From reflexivity of ($\leq$) and transitivity of ($\sqsubseteq$) we then have that both $A+B \sqsubseteq A$ and $C+D \sqsubseteq C$, and thus both $C$ and $D$ must be zero since ($\sqsubseteq$) cannot decrease the overall weight of a hyper-distribution. From this we have that $A \preceq C$ and $C \preceq A$, hence $A = C$ by antisymmetry of ($\preceq$).

The definition of program refinement is the pointwise extension of the above, that is

**Definition 7.4** *Secure Program Refinement*   Let $S, I$ be programs' meanings of type $SS \to \mathbb{D}SS$. We say that $S \sqsubseteq I$ just when for all initial states $s{:}SS$ we have $S.s \sqsubseteq I.s$ according to Def. 7.3.   □

## 7.5  Least fixed points in $SS \to \mathbb{D}SS$: getting around incompleteness

The normal approach to fixed-point semantics for loops would be to show that a loop defines a ($\sqsubseteq$)-continuous functional $\mathcal{L}$ over the program space $SS \to \mathbb{D}SS$, and then to take the ($\sqsubseteq$)-supremum of the chain $\text{II} \sqsubseteq \mathcal{L}.\text{II} \sqsubseteq \mathcal{L}^2.\text{II} \cdots$ where II is the least program, the one producing the output hyper of zero weight for all inputs.

Here instead we show that a loop defines a ($\leq$)-continuous functional $\mathcal{L}$, and then take the ($\leq$)-supremum of the chain $\text{II} \leq \mathcal{L}.\text{II} \leq \mathcal{L}^2.\text{II} \cdots$. Its well definedness follows from Lem. 7.1; its relevance is justified by the following lemma.

**Lemma 7.2** *Equivalence of fixed points*   Let partial orders ($\leq$) and ($\sqsubseteq$) be defined over some space $\mathcal{X}$, and let $\mathcal{L}$ be an endofunction on $\mathcal{X}$. Suppose further that ($\leq$) $\subseteq$ ($\sqsubseteq$), that is that ($\leq$) implies ($\sqsubseteq$).

If a ($\leq$)-least (resp. greatest) fixed point of $\mathcal{L}$ exists, then also a ($\sqsubseteq$)-least (resp. greatest) fixed point of $\mathcal{L}$ exists, and in fact they are equal.
*Proof:*   Let $x$ be the ($\leq$)-least fixed-point of $\mathcal{L}$. Then for any (other) fixed-point $x'$ of $\mathcal{L}$ we have $x \leq x'$ and so –by assumption– also $x \sqsubseteq x'$. Thus $x$ is a ($\sqsubseteq$)-lower-bound for all fixed points; but it is a fixed point itself. Therefore it is the ($\sqsubseteq$)-least fixed point as well. (The same argument holds for greatest.)   □

In the next section we will introduce our language to express and reason about secure programming, extending our previous work with iteration. We use Def. 7.2 for the semantics for loops, relying on Lem. 7.2 to ensure that it is also well-defined as a least fixed point in the security order; we do, of course, need to show that the assumption of ($\leq$)-continuity is satisfied by the semantic definitions we give. In the conclusion we shall return to the question of ($\sqsubseteq$)-limits more generally, i.e. those which are not restricted to ($\leq$)-limits.

## 8  Programming language

Having tied-down the details of our semantic space, we can now give our programs' denotations via structural induction; however there are two potential sources of complexity in what we present. The first, conceptual, is the two-level structure that we motivated in the sections above, the partial distributions that themselves are taken over *other* conditional, or sometimes even *a posteriori* distributions.

The second is notational: standard constructions like conditionals and push-forward are now generated by program fragments that, as a rule, are expressions over free variables (i.e. the variables of the program) rather than (pure) mathematical functions themselves. This leads to uncomfortable expositions like "$\Pr(D|E)$ where distribution $D(x)$ is given by $\cdots x \cdots$ and predicate $E(x)$ holds just when $\cdots x \cdots$". Although these are easy to understand (being well established notations), they are hard to manipulate algebraically in specific cases where $D, E$ are determined by some computer program.

We now introduce specialised notation to streamline our semantic definitions.

### 8.1  Distribution comprehensions

Recall that the *support* $\lceil \delta \rceil$ of distribution $\delta: \mathbb{D}\mathcal{X}$ is those elements $x: \mathcal{X}$ with $\delta.x \neq 0$; naturally for $\delta: \mathbb{D}\mathcal{X}$ we have $\lceil \delta \rceil \subseteq \mathcal{X}$. The *weight* of $\delta$ is written $\sum \delta$, defined $\delta.\mathcal{X}$ so that full distributions have weight 1. Distributions can be scaled and summed according to the usual pointwise extension of multiplication and addition to real-valued functions, provided the outcomes are again distributions.

Given a non-empty finite set $\mathcal{X}$ we write $\lfloor \mathcal{X} \rfloor$ for the uniform distribution over $\mathcal{X}$, that is the uniform distribution $\delta: \mathbb{D}\mathcal{X}$ such that $\lceil \delta \rceil = \mathcal{X}$.

#### 8.1.1  Enumerated distributions and expected values

These are notations for enumerated distributions, i.e. those in which the support is explicitly listed (cf. set enumerations that list a set's elements):

– **empty** The empty, or zero subdistribution has empty support and assigns probability zero to all elements: we write it $\{\!\{\,\}\!\}$.

– **multiple** We write $\{\!\{x^{@p}, y^{@q}, \cdots, z^{@r}\}\!\}$ for the distribution assigning probabilities $p, q, \cdots, r$ to elements $x, y, \cdots, z$ respectively, with $p + q + \cdots + r \leq 1$. Provided $p, q, \cdots, r > 0$, the support is therefore the set $\{x, y, \cdots, z\}$.

- **point** The distribution concentrated on a single element $x$ is written $\{\!|x|\!\}$, i.e. abbreviating $\{\!|x^{@1}|\!\}$ whose support is $\{x\}$.

- **uniform** When explicit probabilities are omitted they are implicitly uniform: thus $\{\!|x, y, z|\!\}$ is $\{\!|x^{@\frac{1}{3}}, y^{@\frac{1}{3}}, z^{@\frac{1}{3}}|\!\}$.

- **binary, and distributed uniform** For a two-element distribution we write $x_p \oplus y$ for $\{\!|x^{@p}, y^{@1-p}|\!\}$, and in the uniform case we can write $x \oplus y \oplus \cdots \oplus z$ for $\{\!|x, y, \cdots, z|\!\}$.

For expected values of random variables that are written as expressions, we have

- **expected value** We write $(\odot d{:}\delta \bullet E)$ for the *expected value* $\sum_{d:\lceil\delta\rceil}(\delta.d \times E)$ of expression $E$, interpreted as a random variable in $d$, over distribution $\delta$.

    If $E$ is Boolean, then it is taken to be 1 if $E$ holds and 0 otherwise, so that the expected value is then just the combined probability in $\delta$ of all elements $d$ satisfying $E$. If necessary for clarity we will write $[E]$ to indicate $E$'s conversion from Boolean to $0, 1$; when possible, however, we omit it (to reduce proliferation of brackets).

### 8.1.2 Distribution comprehensions, conditioning and *a posteriori* values

As for set comprehensions, with distribution comprehensions we describe a distribution by giving a rule for forming it, i.e. its supporting elements and the probabilities they have. Here are the common cases:

- **map, push-forward** When $f$ in §3.4 is given as an expression $E$ of type $\mathcal{Y}$, with free variable $x$ say, then for the push-forward distribution $\mathsf{map}.f.\delta$ we write the comprehension $\{\!|x{:}\delta \bullet E|\!\}$ where for $y{:}\mathcal{Y}$ we define

$$\{\!|x{:}\delta \bullet E|\!\}.y \quad := \quad (\odot x{:}\delta \bullet E{=}y) \ .$$

    Recall from above that the Boolean value $E{=}y$ is to be converted implicitly to $0, 1$ in this case.

- **conditional distribution** Given a distribution $\delta{:}\mathbb{D}\mathcal{X}$ and a Boolean expression $R$ in free variable $x$, we write $\{\!|x{:}\delta \mid R|\!\}$ for the distribution obtained by conditioning $\delta$ on the set (the event) that $R$ represents as a predicate in $x$. Thus for $x'{:}\mathcal{X}$ we have

$$\{\!|x{:}\delta \mid R|\!\}.x' \quad := \quad \delta.x' \times [R'] \ / \ (\odot x{:}\delta \bullet [R]) \ , \tag{3}$$

    where $R'$ is $R$ with $x$ replaced by $x'$ and here, for clarity, with $[\cdot]$ we make the conversions to $0, 1$ explicit.

- ***a posteriori* values** Finally, for *Bayesian belief revision* suppose $\delta$ is an *a priori* distribution over some $\mathcal{X}$ and let expression $R$ (not Boolean) in free variable $x$ in $D$ be the probability of a certain observable result if that $x$ is chosen. Then $\{\!|x\!:\!\delta \mid R|\!\}$ is the *a posteriori* distribution (revising $\delta$) when that result actually occurs. The definition is as for (3) immediately above, but using just $R$ rather than $[R]$.

Note that $R$ can be scaled without affecting the value of this expression, so wlog it can be made one-summing as $x$ varies: this makes it easier to interpret as a probabilistic outcome that triggers Bayesian belief revision.

– **general distribution comprehension** We can combine all the above possibilities by writing $\{\!|x\!:\!\delta \mid R \bullet E|\!\}$, for distribution $\delta$, real expression $R$ (in $x$) and expression $E$ (also in $x$) to mean

$$(\odot x\!:\!\delta \bullet R \times \{\!|E|\!\}) \:/\: (\odot x\!:\!\delta \bullet R) \qquad (4)$$

where, first, an expected value is formed in the numerator by scaling and adding point-distribution $\{\!|E|\!\}$ as a real-valued function: this gives another (sub-)distribution. The scalar denominator then conditions on $R$.

A missing $E$ is implicitly $x$ itself. If $R$ is omitted, then $(R\times)$ is removed from the numerator, and the denominator is removed altogether. (When $\delta$ is a full distribution, this happens automatically by assuming a missing $R$ to be 1, or equivalently Boolean *true*.)

As a concrete example we recall the puzzle

In families with two children of equally and independently distributed gender, if one child is a boy what is the chance that the other is too?

Encoding *boy,girl* as Booleans $\mathsf{T},\mathsf{F}$ we write $\{\!|x,y\!:\!\mathsf{T}\oplus\mathsf{F} \mid x \vee y \bullet x \wedge y|\!\}$ for the distribution of the pushed-forward function *both boys* $(x \wedge y)$ over the *iid gender joint-distribution* of the two children $(x, y\!:\!\mathsf{T}\oplus\mathsf{F})$ conditioned on the event *at least one boy* $(x \vee y)$. It works out as

$$\begin{aligned}
& \{\!|x,y\!:\!\mathsf{T}\oplus\mathsf{F} \mid x \vee y \bullet x \wedge y|\!\} \\
=\ & (\odot x, y\!:\!\mathsf{T}\oplus\mathsf{F} \bullet [x \vee y] \times \{\!|x \wedge y|\!\}) \:/\: (\odot x, y\!:\!\mathsf{T}\oplus\mathsf{F} \bullet [x \vee y]) \\
=\ & (1 \times \{\!|\mathsf{T}|\!\}/4 + 1 \times \{\!|\mathsf{F}|\!\}/4 + 1 \times \{\!|\mathsf{F}|\!\}/4 + 0 \times \{\!|\mathsf{F}|\!\}/4) \:/\: (3/4) \\
=\ & \{\!|\mathsf{T}^{@\frac{1}{4}}, \mathsf{F}^{@\frac{1}{2}}|\!\} \:/\: (3/4) \\
=\ & \mathsf{T}\ _{1/3}\!\oplus\ \mathsf{F}\ ,
\end{aligned}$$

that is (as we know) that the *a posteriori* probability of "both boys" is 1/3. This is the kind of calculation that specific programs' semantics generate.

## 8.2 Program semantics for the *HMM* core: revelations

We recall from §3.1 that an *HMM* is determined by two stochastic kernels (matrices) $\boldsymbol{T}, \boldsymbol{E}$. In programming terms the $\boldsymbol{T}$ represents a probabilistic assignment to our hidden variable h; we deal with that at *Choose prob. hidden* in §8.3 below.

The $\boldsymbol{E}$ on the other hand releases information (about h) in what we call a "revelation" — observables our attacker can see [29]. It has two forms, the second a generalisation of the first.

In the two definitions below, and further, we write $E.\mathsf{v}.\mathsf{h}$ as an *expression* in which program variables $\mathsf{v}, \mathsf{h}$ might occur free. The same convention applies to $D, G, p$ for distributions, (Boolean) guards and probabilities resp.

| Program type | Program text $P$ | Semantics $[\![P]\!].(v, \delta)$ |
|---|---|---|
| *Reveal value* | **reveal** $E.\mathsf{v}.\mathsf{h}$ | $\{\!|h\!:\!\delta \bullet (v, \{\!|h'\!:\!\delta \mid E.v.h'\!=\!E.v.h|\!\})|\!\}$ |

Expression $E.\mathsf{v}.\mathsf{h}$ takes its value in some type $\mathcal{X}$ representing observations an attacker can make. The command reveals a value $x$ depending on $\mathsf{v}, \mathsf{h}$. Neither $\mathsf{v}$ nor $\mathsf{h}$ is changed by this; but the outgoing distribution of the hidden $\mathsf{h}$ is conditioned on the basis of the $x$ revealed. Note that $x$ is not stored; but because of perfect recall an attacker can remember it.

| | | |
|---|---|---|
| *Reveal choice* | **reveal** $D.\mathsf{v}.\mathsf{h}$ | $\{\!|h\!:\!\delta; x\!:\!D.v.h \bullet (v, \{\!|h\!:\!\delta \mid D.v.h.x|\!\})|\!\}$ |

Expression $D.\mathsf{v}.\mathsf{h}$ is now more generally of type $\mathbb{D}\mathcal{X}$, so that for $x\!:\!\mathcal{X}$ we have $D.\mathsf{v}.\mathsf{h}.x$ as a probability. The command calculates that distribution, and then chooses some value $x$ according to those probablities; that value $x$ is then revealed. As before, variables $\mathsf{v}, \mathsf{h}$ are not changed; but the distribution of $\mathsf{h}$ is conditioned on the fact that $x$ was revealed.

*Reveal value* is the special case **reveal** $\{\!|E.\mathsf{v}.\mathsf{h}|\!\}$ of *Reveal choice*.

## 8.3 Semantics of syntactically atomic commands

Syntactically atomic commands are regarded as *semantically* atomic in the sense that the only information they leak is what the final value of the visible $\mathsf{v}$ allows to be deduced about the final value of $\mathsf{h}$ with knowledge of the program text. Thus for example $\mathsf{v}\!:=\!\mathsf{h}$ leaks everything about $\mathsf{h}$, since $\mathsf{v}$'s final value is evidently the same as $\mathsf{h}$'s; yet $\mathsf{v}\!:=\!0\!\times\!\mathsf{h}$ reveals nothing, even though at some point in an internal register the value of $\mathsf{h}$ might have been accessible. In this sense the syntactic atoms are the atoms of observation also: within them neither perfect recall nor implicit flow make sense.

We determine the semantics of these atomic commands systematically. Using "classical," i.e. without-noninterference probabilistic sequential semantics [25, etc.] gives a straightforward meaning to atomic commands' actions on a state space $SS$ as functions of type $SS \to \mathbb{D}SS$ taking an initial distribution to a (sub-)distribution of final states. If we abstract from noninterference properties by considering $\mathsf{v}$ to be hidden (as well as $\mathsf{h}$), and set $SS\!:=\!\mathcal{V}\!\times\!\mathcal{H}$ then we have a ready-made classical semantics for the syntactic atoms we are dealing with here.

The initial "state" will be a pair $(v, \delta)$ in $\mathcal{V}\!\times\!\mathbb{D}\mathcal{H}$. We therefore reuse "$Q$" from Def. 7.1 to express this as the joint distribution $\{\!|v|\!\}\!\times\!\delta$ of type $\mathbb{D}(\mathcal{V}\!\times\!\mathcal{H})$, that is $\mathbb{D}SS$. To apply a command with semantics of type $SS \to \mathbb{D}SS$ to that, we use lifting (§3.4) so that the result of this classical interpretation is again of type $\mathbb{D}(\mathcal{V}\!\times\!\mathcal{H})$, and we convert this back to the noninterference output-type $\mathbb{D}(\mathcal{V}\!\times\!\mathbb{D}\mathcal{H})$ by analogy with "revealing $\mathsf{v}$" according to the semantics above — since knowledge of $\mathsf{v}$'s final value is all that escapes an atomic command. Fol-lowing *Reveal value* from above, we define

$$\mathsf{rv}.\Delta \quad := \quad \{\!|(v,h)\!:\!\Delta \bullet (v, \{\!|(v',h')\!:\!\Delta) \mid v{=}v'\}\!|)|\!\} \ .^{11} \qquad (5)$$

The result of the procedure above –convert incoming $\mathcal{V}{\times}\mathbb{D}\mathcal{H}$ to $\mathbb{D}(\mathcal{V}{\times}\mathcal{H})$, then apply lifted classical semantics; then apply rv to the result– is summarised below. Observe that neither program **abort**, nor assertions are necessarily useful for writing specific programs, but our focus is on reasoning *about* programs, in particular algebraically, and for that these commands play a prominent role.

Program type[12]      Program text $P$      Semantics $[\![P]\!].(v,\delta)$

*Least element*      **abort**      $\{\!|\}\!|$

This is the program that simply fails to terminate: for every input it produces the empty subdistribution as output. In our refinement order, as a specification it allows all possible implementations (i.e. that **abort** $\sqsubseteq S$ for all $S$) — essentially playing the role of "0" in arithmetic.

*Identity*      **skip**      $\{\!|(v,\delta)|\!\}$

The "do nothing" command simply converts its input to a point-hyper on output, i.e. reproduces its input with probability one.

*Assertion*      $\{p.\mathsf{v}.\mathsf{h}\}$      $\{\!|(v, \{\!|h\!:\!\delta \mid p.v.h|\!\}) @(\odot h'\!:\!\delta \bullet p.v.h') |\!\}$

An assertion gives directly in $p.\mathsf{v}.\mathsf{h}$ the a probability of the command's termination. With probability $1{-}p$ the assertion behaves as **abort**.

When with probability $p$ it *does* terminate, however, it conditions the hidden value's distribution $\delta$ on the fact it did so: that is $\delta$ is revised to reflect that the abort did not occur. The visible variable v is unaffected in this case.

*Assign to visible*      $\mathsf{v}{:=}E.\mathsf{v}.\mathsf{h}$      $\{\!| h\!:\!\delta \bullet (E.v.h, \{\!|h'\!:\!\delta \mid E.v.h'{=}E.v.h|\!\}) |\!\}$

The command's effect is to assign the *rhs*-value to v *but also* to condition the hidden distribution on the fact that h can produce the value observed to have been put into v.

*Assign to hidden*      $\mathsf{h}{:=}E.\mathsf{v}.\mathsf{h}$      $\{\!| (v, \{\!|h\!:\!\delta \bullet E.v.h|\!\}) |\!\}$

The command does not change v, but maps the hidden incoming distribution of h through $E.\mathsf{v}$ considered as a function of (incoming) h to produce the resulting distribution on (outgoing) h.

*Choose prob. visible*      $\mathsf{v}{:}{\in} D.\mathsf{v}.\mathsf{h}$

$$\{\!| v'\!:\!(\odot h\!:\!\delta \bullet D.v.h) \bullet (v', \{\!|h'\!:\!\delta \mid D.v.h'.v'|\!\}) |\!\}$$

Expression $D.\mathsf{v}.\mathsf{h}$ is a *distribution* on $\mathcal{V}$, and the choice of v's new value is made according to it. It generalises *Assign to visible*, since the latter can be written $\mathsf{v}{:}{\in}\{\!|E.\mathsf{v}.\mathsf{h}|\!\}$.

*Choose prob. hidden*      $\mathsf{h}{:}{\in} D.\mathsf{v}.\mathsf{h}$      $\{\!| (v, (\odot h\!:\!\delta \bullet D.v.h)) |\!\}$

---

[11] We justify (5) informally by noting that it's what results from replacing hidden h in the *rhs* of *Reveal value* by the hidden pair $(\mathsf{v},\mathsf{h})$ and considering the expression $E.\mathsf{v}.\mathsf{h}$ to be simply v.

Expression $D.\mathsf{v}.\mathsf{h}$ is now a distribution on $\mathcal{H}$, and the choice of $\mathsf{h}$'s new value is made according to it. It generalises *Assign to hidden*, since the latter can be written $\mathsf{h}{:}\in \{\!|E.\mathsf{v}.\mathsf{h}|\!\}$.

As a syntactic convenience, when we are using the more general "choose" form of either command but the *rhs*'s distribution is written out using ($\oplus$) rather than as a $\{\!|\,|\!\}$-style comprehension, we use the conventional assignment symbol $(:=)$ so that e.g. we can write $\mathsf{v}{:=}\mathsf{T}\oplus\mathsf{F}$ for flipping a fair Boolean coin.

As an example of the algebraic utility of *Assertion*, we note that distinguished commands **abort** and **skip** are special cases of assertions, so that **skip** = $\{\mathsf{T}\}$ and **abort** = $\{\mathsf{F}\}$. Further, the semantics of *Reveal choice* can be given more compactly –assuming $D.\mathsf{v}.\mathsf{h}$ has type $\mathbb{D}\mathcal{X}$– as

$$[\![\mathbf{reveal}\ D.\mathsf{v}.\mathsf{h}]\!].(v,\delta) \quad = \quad (\ \sum x{:}\mathcal{X}\bullet [\![\{D.\mathsf{v}.\mathsf{h}.x\}]\!].(v,\delta)\ )\ . \qquad (6)$$

That formulation makes it easy to reason about revelations in terms of more primitive commands. We also have that assignments to visible variables that may depend on $\mathsf{h}$ may be represented more simply in terms of those that do not:

$$[\![\mathsf{v}{:}\in D.\mathsf{v}.\mathsf{h}]\!].(v,\delta) \quad = \quad (\ \sum v'{:}\mathcal{V}\bullet [\![\{D.\mathsf{v}.\mathsf{h}.v'\};\mathsf{v}{:=}v']\!]\ )\ . \qquad (7)$$

As we will see in the next section, assertions also play an important role in the specification of probabilistic choice and conditionals.

## 8.4 Semantics of compound commands: implicit flow

Compound commands are in fact the simplest to define, since they are treated almost as they would be for classical semantics. The only adjustment is to insert conditioning assertions on program branch-points to enforce *implicit flow*, that is that information escapes by observation of the outcome of conditionals.

| Program type | Program text $P$ | Semantics $[\![P]\!].(v,\delta)$ |
|---|---|---|
| *Composition* | $P_1;P_2$ | $[\![P_2]\!]^*.([\![P_1]\!].(v,\delta))$ |

    Sequential composition is interpreted as Kleisli composition (§3.4).

*General prob. choice*      $P_L {}_{p.\mathsf{v}.\mathsf{h}}\!\oplus P_R$
$$[\![\{p.\mathsf{v}.\mathsf{h}\};P_L]\!].(v,\delta) + [\![\{1{-}p.\mathsf{v}.\mathsf{h}\};P_R]\!].(v,\delta)$$

    Expression $p.\mathsf{v}.\mathsf{h}$ is evaluated to a probability of the command's taking its left branch; otherwise it takes the right. The attacker can observe which branch was taken: this is reflected in the conditioning assertions at the beginning of each branch.

*Conditional choice*      **if** $G.\mathsf{v}.\mathsf{h}$ **then** $P_T$ **else** $P_F$ **fi**

---

[12]The most general form of atomic assignment is the *Simultaneous choice* mentioned earlier, whose semantics can be deduced as for the others from its classical behaviour. Since it is seldom needed, however, we omit its definition for brevity.

$$[\![\{G.\mathsf{v}.\mathsf{h}\}; P_T]\!].(v,\delta) + [\![\{\neg G.\mathsf{v}.\mathsf{h}\}; P_F]\!].(v,\delta)$$

This is a specialisation of the previous *General probabilistic choice* to the case where the probability is always either 1 (go left) or 0 (go right). Again the conditioning assertions guard each branch.

*Iteration*                    **while** $p.\mathsf{v}.\mathsf{h}$ **do** $P$ **od**    the ($\leq$)-least fixed point of $\mathcal{L}$, applied to $(v,\delta)$, where $\mathcal{L}$ is the unique endofunction on the space $SS\to\mathbb{D}SS$ of programs' meanings such that for any program $L$ we have $[\![P; L \;_{p.\mathsf{v}.\mathsf{h}}\oplus \mathbf{skip}]\!] = \mathcal{L}.[\![L]\!]$.

As for *Conditional choice*, the loop guard is a probability determined by the program variables $\mathsf{v},\mathsf{h}$, with as a special case Booleans $\mathsf{T},\mathsf{F}$ interpreted as 1 (enter the loop) or 0 (terminate the loop).

For iteration we are taking the usual least-fixed-point approach except, for the reasons explained above, we use a special *termination order* ($\leq$) for the chain of iterates. For this we need the (usual) technical results of continuity of our program contexts.

**Lemma 8.1** *Continuity of program contexts*    Any context $\mathcal{C}(\cdot)$, constructed in the programming language above, satisfies $\mathcal{C}(\bigvee_i P_i) = \bigvee_i \mathcal{C}(P_i)$ for non-empty ($\leq$)-chains $\bigvee_i P_i$.

*Proof:*    Because the termination order is so simple (unlike the entropy order), being essentially pointwise less-than-or-equals, this result follows easily from linearity of the Kleisli-composition (essentially lifting) used in the definition of sequential composition.                                                         □

Importantly, each our compound operators are monotonic with respect to their arguments and the secure refinement order ($\sqsubseteq$), meaning that we may reason compositionally about the correctness of programs.

**Theorem 8.1** *Monotonicity of compound commands*    Each of the commands listed above are monotonic with respect to their program arguments and the refinement order.                □

## 8.5   Local- and multiple variables; hidden correlations

To this point we have had just two variables, visible $\mathsf{v}$ and hidden $\mathsf{h}$, and have been assuming for simplicity that they are all the variables in the program. In practice however each of $\mathcal{V},\mathcal{H}$ will each comprise many variables, represented in the usual Cartesian way. Thus if we have variables $\mathsf{a}\colon\mathcal{A}, \mathsf{b}\colon\mathcal{B}, \mathsf{c}\colon\mathcal{C}, \mathsf{d}\colon\mathcal{D}$ with the first two $\mathsf{a},\mathsf{b}$ visible and the last two $\mathsf{c},\mathsf{d}$ hidden, then $\mathcal{V}$ is $\mathcal{A}\times\mathcal{B}$ and $\mathcal{H}$ is $\mathcal{C}\times\mathcal{D}$ so that the state-space is $\mathcal{A}\times\mathcal{B}\times\mathbb{D}(\mathcal{C}\times\mathcal{D})$. Assignments and projections are handled as normal.

Thus we allow local variables, both visible and hidden, which extend the state as described above: within the scope of a visible local-variable declaration $|[[$ **vis** x: $\mathcal{X} \cdots \,]]|$, the $\mathcal{V}_{\text{local}}$ used is $\mathcal{X} \times \mathcal{V}_{\text{global}}$. Hidden variables are similar.[13]

Note however that because for simplicity we have been assuming that v, h are in fact *all* the variables in the program, i.e. that they stand for vectors of variables implicitly, our semantics above establishes the equality of the two fragments v:= h; v, h:= 0, 0 and v, h:= 0, 0, reflecting our deliberate concentration on h's *final* value [36, 37] in order to extend conventional refinement [33, 5] that does the same. In this case h's initial value's being revealed on the left has no bearing on our knowledge or ignorance of its final value and so does not introduce a difference in meaning between the two fragments shown.

If however there are other hidden variables, not mentioned but still in scope as might happen within a local block or within the context of extra declarations, then our semantics must be slightly more general, in particular recognising that the v or h appearing on the left of an assignment is just one component of a vector of visible resp. hidden variables.

Technically this is handled by extending our hidden distribution to type $\mathbb{D}\mathcal{H}^2$, which tracks correlations with initial values. For simplicity we do not do that here, since in fact any program in which hiddens are not assigned-to (as in our examples and case studies) can be treated with the simpler $\mathbb{D}\mathcal{H}$-style semantics.

## 9  Algebra of HMM-style programs

The programming language introduced in §8, interpreted over the hyper-based semantics, admits a program algebra allowing the proof of general refinements between programs. In this section we present some of the foundational laws of this program algebra, which are then illustrated in §9.8 and §10, via an example based on password guessing.

### 9.1  General principles and scoping laws; referential transparency

As for classical programs, it is possible to replace expressions by other expressions of equal value in context so that, for example, referential transparency gives

$$\textsf{v}:= E.\textsf{v}.\textsf{h}; \{\textsf{T}\} \quad = \quad \textsf{v}:= E.\textsf{v}.\textsf{h}; \{\textsf{v}=E.\textsf{v}.\textsf{h}\} \ .$$

It is also possible to move program fragments in and out of local scopes provided variable bindings are respected. Since empty scopes are equivalent to **skip**, i.e.

$$\textbf{skip} \quad = \quad |[[\ \textbf{vis}\ \textsf{v}':\mathcal{V}\ ]]| \quad = \quad |[[\ \textbf{hid}\ \textsf{h}':\mathcal{H}\ ]]| \ , \tag{8}$$

---
[13]Implicitly local variables are assumed to be initialised by a uniform choice over their finite state space. In our examples however, we always initialize local variables explicitly, to avoid confusion.

it is possible to introduce fresh variables of any constant type. We may also introduce assignments to scope-terminated variables as long as they do not reveal information about the hidden state:

$$|[ \textbf{vis } \mathsf{v}':\mathcal{V}; \ \cdots \ ]| \quad = \quad |[ \textbf{vis } \mathsf{v}':\mathcal{V}; \ \cdots \ ; \mathsf{v}'{:}{\in} D.\mathsf{v}.\mathsf{v}' \ ]| \ , \qquad (9)$$

$$|[ \textbf{hid } \mathsf{h}':\mathcal{H}; \ \cdots \ ]| \quad = \quad |[ \textbf{hid } \mathsf{h}':\mathcal{H}; \ \cdots \ ; \mathsf{h}'{:}{\in} D.\mathsf{v}.\mathsf{h}' \ ]| \ . \quad (10)$$

As an example of the interaction of local scopes and visibility we have

$$
\begin{aligned}
&\ [\![ \textbf{ reveal } D.\mathsf{v}.\mathsf{h} \ ]\!] \\
=\ &\ (\ \textstyle\sum v'{:}\mathcal{X} \bullet [\![ \{D.\mathsf{v}.\mathsf{h}.v'\} ]\!] \ ) \qquad \text{"represent revelation using assertions (6)"} \\
=\ &\ (\ \textstyle\sum v'{:}\mathcal{X} \bullet [\![\ |[ \textbf{vis } \mathsf{v}'{:}\mathcal{X}; \{D.\mathsf{v}.\mathsf{h}.v'\}; \mathsf{v}'{:=}v' \ ]| \ ]\!] \ ) \ \text{"introduce fresh variable} \\
&\qquad\qquad\qquad\qquad\qquad\qquad\qquad\qquad\qquad\text{terminated by a secure assignment"} \\
=\ &\ [\![\ |[ \textbf{vis } \mathsf{v}'{:}\mathcal{X};\ \mathsf{v}'{:}{\in} D.\mathsf{v}.\mathsf{h} \ ]| \ ]\!] \ , \qquad\qquad \text{"shift scope and represent} \\
&\qquad\qquad\qquad\qquad\qquad\qquad\text{visible assignment using assertions (7)"}
\end{aligned}
$$

i.e. that a revelation is effectively an assignment to a temporary visible variable: because of perfect recall, the revealed value is not forgotten; but because the temporary variable is declared within a block, it is effectively erased.

## 9.2 Assertions

We present here some basic properties of assertions that will be used to justify algebraic laws for more complex statements such as revelations and probabilistic choices. First, we have that assertions satisfy the following equivalence,

$$\{p_1.\mathsf{v}.\mathsf{h}\}; \{p_2.\mathsf{v}.\mathsf{h}\} \quad = \quad \{p_1.\mathsf{v}.\mathsf{h} \times p_2.\mathsf{v}.\mathsf{h}\} \quad = \quad \{p_2.\mathsf{v}.\mathsf{h}\}; \{p_1.\mathsf{v}.\mathsf{h}\} \quad (11)$$

and are thus commutative under sequential composition. Constant assertions also commute over arbitrary programs, so that

$$\{p\}; S \quad = \quad S; \{p\} \ . \qquad (12)$$

Since assertions referring to h may condition the hidden state, from the definition of secure refinement (Def. 7.3) we have

$$(\ \textstyle\sum n \bullet [\![\{p_n.\mathsf{v}.\mathsf{h}\}]\!]\ ) \quad \sqsubseteq \quad [\![\{\ \textstyle\sum n \bullet p_n.\mathsf{v}.\mathsf{h}\ \}]\!] \ , \qquad (13)$$

for $(\ \sum n \bullet p_n.\mathsf{v}.\mathsf{h}\ ) \leq 1$. Using this we can calculate that $\textbf{skip}\ _{p.\mathsf{v}.\mathsf{h}}\oplus \textbf{skip} \sqsubseteq \textbf{skip}$ since from implicit flow the *lhs* reveals $p.\mathsf{v}.\mathsf{h}$ but the *rhs* reveals nothing.

On the other hand, additions of assertions that refer only to the *visible* state reveal nothing, and thus (13) can be strengthened to equality, giving

$$(\ \textstyle\sum n \bullet [\![\{p_n.\mathsf{v}\}]\!]\ ) \quad = \quad [\![\{\ \textstyle\sum n \bullet p_n.\mathsf{v}\ \}]\!] \ , \qquad (14)$$

whence $\textbf{skip}\ _{p.\mathsf{v}}\oplus \textbf{skip} = \textbf{skip}$.

Using this algebra of assertions for Booleans $G_{\{1,2\}}.\mathsf{v}.\mathsf{h}$ we have

$$\{G_1.\mathsf{v}.\mathsf{h}\}; \{G_2.\mathsf{v}.\mathsf{h}\} \quad = \quad \{G_1.\mathsf{v}.\mathsf{h} \wedge G_2.\mathsf{v}.\mathsf{h}\}, \qquad (15)$$

and so the *Boolean assertions* are idempotent, that is $\{G.\mathsf{v}.\mathsf{h}\};\{G.\mathsf{v}.\mathsf{h}\} = \{G.\mathsf{v}.\mathsf{h}\}$, and complements under composition so that $\{G.\mathsf{v}.\mathsf{h}\};\{\neg G.\mathsf{v}.\mathsf{h}\} = \mathbf{abort}$. When all Boolean $G_n.\mathsf{v}.\mathsf{h}$ are disjoint we also have

$$( \sum n\colon [1..N] \bullet \{G_n.\mathsf{v}.\mathsf{h}\} ) \quad \sqsubseteq \quad \{ \bigvee n\colon [1..N] \bullet G_n.\mathsf{v}.\mathsf{h} \} , \qquad (16)$$
$$( \sum n\colon [1..N] \bullet \{G_n.\mathsf{v}\} ) \quad = \quad \{ \bigvee n\colon [1..N] \bullet G_n.\mathsf{v} \} . \qquad (17)$$

## 9.3 Basic laws for revelations

A single **reveal** releases information but changes no variable. Using refinement we can with **reveal** $D_1 \sqsubseteq$ **reveal** $D_2$ express that revealing $D_2$ leaks no more information than revealing $D_1$ would have. The refinement between programs means this statement applies for *any* incoming distribution.

We write **reveal** $(E_1, E_2)$ for the release of two pieces of information, one defined by expression $E_1$ and the other defined by expression $E_2$. For example **reveal** $(\mathsf{h}\,\mathbf{mod}\,2, \mathsf{h}\,\mathbf{mod}\,3)$ releases information about both $\mathsf{h}$'s divisibility by 2 and 3: this is more informative than releasing just one, giving the refinement

$$\mathbf{reveal}\ (\mathsf{h}\,\mathbf{mod}\,2, \mathsf{h}\,\mathbf{mod}\,3) \quad \sqsubseteq \quad \mathbf{reveal}\ \ \mathsf{h}\,\mathbf{mod}\,2 . \qquad (18)$$

As we shall see, this and a number of other laws can be derived from a single general refinement rule which effectively states that any released information can be concealed somewhat by distributing it stochastically.

**Lemma 9.1** *Basic reveal refinement*  Let $D.\mathsf{v}.\mathsf{h}$ be a distribution over some $\mathcal{X}$ and $F$ be a stochastic matrix (which can depend on $\mathsf{v}$) giving for each element of $\mathcal{X}$ a distribution over some other type $\mathcal{Y}$. Then we have

$$\mathbf{reveal}\ D.\mathsf{v}.\mathsf{h} \quad \sqsubseteq \quad \mathbf{reveal}\ D.\mathsf{v}.\mathsf{h} \otimes F.\mathsf{v} ,$$

where $(\otimes)$ is defined by $(D.\mathsf{v}.\mathsf{h} \otimes F.\mathsf{v}).y := ( \sum x\colon \mathcal{X} \bullet D.\mathsf{v}.\mathsf{h}.x \times F.\mathsf{v}.x.y )$. [14]
*Proof:*  We reason as follows:

$$\begin{aligned}
&\llbracket \mathbf{reveal}\ D.\mathsf{v}.\mathsf{h} \rrbracket \\
=\ & ( \sum x\colon \mathcal{X} \bullet \llbracket \{D.\mathsf{v}.\mathsf{h}.x\} \rrbracket ) && \text{``define revelation using assertions (6)''} \\
\\
=\ & ( \sum x\colon \mathcal{X} \bullet \llbracket \{D.\mathsf{v}.\mathsf{h}.x\}; \{( \sum y\colon \mathcal{Y} \bullet F.\mathsf{v}.x.y )\} \rrbracket ) && \text{``distribution } F.\mathsf{v} \text{ is full;} \\
& && \{1\} = \mathbf{skip} \text{ is unit of composition (33)''} \\
\\
=\ & ( \sum x, y\colon \mathcal{X}, \mathcal{Y} \bullet \llbracket \{D.\mathsf{v}.\mathsf{h}.x\}; \{F.\mathsf{v}.x.y\} \rrbracket ) && \text{``(14); Kleisli composition} \\
& && \text{distributes over addition''} \\
\\
=\ & ( \sum x, y\colon \mathcal{X}, \mathcal{Y} \bullet \llbracket \{D.\mathsf{v}.\mathsf{h}.x \times F.\mathsf{v}.x.y\} \rrbracket ) && \text{``(11)''} \\
\sqsubseteq\ & ( \sum y\colon \mathcal{Y} \bullet \llbracket \{( \sum x\colon \mathcal{X} \bullet D.\mathsf{v}.\mathsf{h}.x \times F.\mathsf{v}.x.y )\} \rrbracket ) && \text{``(13)''} \\
=\ & \llbracket \mathbf{reveal}\ D.\mathsf{v}.\mathsf{h} \otimes F.\mathsf{v} \rrbracket . && \text{``define revelation using assertions (6)''}
\end{aligned}$$

---

[14] If $D.\mathsf{v}.\mathsf{h}$ and $F.\mathsf{v}$ are expressed as matrices then $(\otimes)$ is matrix multiplication.

☐

As an example of Lem. 9.1 we suppose h is Boolean, and that we have a revelation behaving as follows. If h is T then T is emitted with probability 1/4 and F with probability 3/4; if h is F then F is emitted unconditionally. We write this **reveal** $D$.h (omitting the .v in this simple case) via the $D$-matrix

$$\begin{array}{cc} \mathsf{T} & \mathsf{F} \end{array} \quad \leftarrow \text{emitted value}$$
$$\begin{array}{c} \mathsf{h}=\mathsf{T} \\ \mathsf{h}=\mathsf{F} \end{array} \begin{pmatrix} 1/4 & 3/4 \\ 0 & 1 \end{pmatrix} \tag{19}$$

Now we can condition on the emitted value, so defining a partition on any incoming state: for example if the incoming state $s$ is $(v, \{\!\!\{\mathsf{T}^{@\frac{1}{2}}, \mathsf{F}^{@\frac{1}{2}}\}\!\!\})$ then $[\![\mathbf{reveal}\ D.\mathsf{h}]\!].s = \{\!\!\{(v, \{\!\!\{\mathsf{T}\}\!\!\})^{@\frac{1}{8}}, (v, \{\!\!\{\mathsf{T}^{@\frac{3}{7}}, \mathsf{F}^{@\frac{4}{7}}\}\!\!\})^{@\frac{7}{8}}\}\!\!\}$ expressing the fact that T is emitted only if h is T thus completely revealing h in this case; however this happens only 1/8 of the time; the remaining 7/8 of the time h is only partly revealed, with the *a posteriori* distribution's being merely F-skewed.

Now suppose that the process is overlaid by another process $F$ (again omitting .v) which obscures the information emitted by **reveal** $D$.h by changing the values stochastically:

$$\begin{array}{cc} \mathsf{T} & \mathsf{F} \end{array} \quad \leftarrow \text{new emitted value}$$
$$\begin{array}{c} \text{emission from } D.\mathsf{h} \text{ was } \mathsf{T} \\ \text{emission from } D.\mathsf{h} \text{ was } \mathsf{F} \end{array} \begin{pmatrix} 1 & 0 \\ 1/2 & 1/2 \end{pmatrix} \tag{20}$$

Overall, the value actually emitted by the combination is determined by the product of the matrices in (19) and (20), that is

$$\begin{pmatrix} 1/4 & 3/4 \\ 0 & 1 \end{pmatrix} \times \begin{pmatrix} 1 & 0 \\ 1/2 & 1/2 \end{pmatrix} = \begin{pmatrix} 5/8 & 3/8 \\ 1/2 & 1/2 \end{pmatrix}$$

which for the chosen incoming distribution gives that $[\![\mathbf{reveal}\ D.\mathsf{h} \otimes F]\!].s$ is $\{\!\!\{(v, \{\!\!\{\mathsf{T}^{@\frac{5}{9}}, \mathsf{F}^{@\frac{4}{9}}\}\!\!\})^{@\frac{9}{16}}, (v, \{\!\!\{\mathsf{T}^{@\frac{3}{7}}, \mathsf{F}^{@\frac{4}{7}}\}\!\!\})^{@\frac{7}{16}}\}\!\!\}$, leaking less than $[\![\mathbf{reveal}\ D.\mathsf{h}]\!].s$.

Now Lem. 9.1 justifies (18) with $F$ as the projection function onto the first component. Other rules can be derived similarly:

**Lemma 9.2** *Simple reveal rules*

$$\begin{array}{rcll} \mathbf{reveal}\ k & = & \mathbf{skip} & (21) \\ \mathbf{reveal}\ D.\mathsf{v}.\mathsf{h} & \sqsubseteq & \mathbf{skip} & (22) \\ \mathbf{reveal}\ \mathsf{h} & \sqsubseteq & \mathbf{reveal}\ D.\mathsf{v}.\mathsf{h} & (23) \\ \mathbf{reveal}\ G.\mathsf{v}.\mathsf{h} & = & \mathbf{reveal}\ \neg G.\mathsf{v}.\mathsf{h} & (24) \\ \mathbf{reveal}\ (E_1.\mathsf{v}.\mathsf{h}, E_2.\mathsf{v}.\mathsf{h}) & \sqsubseteq & \mathbf{reveal}\ E_1.\mathsf{v}.\mathsf{h} & (25) \\ \mathbf{reveal}\ (E.\mathsf{v}.\mathsf{h}, E.\mathsf{v}.\mathsf{h}) & = & \mathbf{reveal}\ E.\mathsf{v}.\mathsf{h} & (26) \end{array}$$

*Proof:* The first is a consequence of the equivalent definition of revelations in terms of assertions and the rest are consequences of it and Lem. 9.1. For example

(22) follows by defining $F.\mathsf{v}$ to be constant; and (23) follows by defining $F.\mathsf{v}$ in Lem. 9.1 to be $D.\mathsf{v}.\mathsf{h}$; (24) follows by defining $F.\mathsf{v}$ to swap the values $\mathsf{T}$ and $\mathsf{F}$. Finally (25,26) follow by defining $F.\mathsf{v}$ to be the projection function. □

With the apparatus so far, the example in §2 could be sketched [15]

$$
\begin{array}{rl}
& \mathsf{v}{:=}\mathsf{h}{\div}2;\ \mathsf{v}{:=}\mathsf{v}{\div}2 \\
= & |\![[\ \mathbf{vis}\ \mathsf{v}';\ \mathsf{v}'{:=}\mathsf{h}{\div}2; \mathsf{v}{:=}\mathsf{v}'{\div}2\ ]\!]| \quad \text{``classical reasoning with visibles and scopes''} \\
= & |\![[\ \mathbf{vis}\ \mathsf{v}';\ \mathsf{v}'{:=}\mathsf{h}{\div}2; \mathsf{v}{:=}(\mathsf{h}{\div}2){\div}2\ ]\!]| \quad\quad \text{``referential transparency §9.1''} \\
= & |\![[\ \mathbf{vis}\ \mathsf{v}';\ \mathsf{v}'{:=}\mathsf{h}{\div}2\ ]\!]|; \mathsf{v}{:=}\mathsf{h}{\div}4 \quad\quad \text{``shrink scope; arithmetic''} \\
= & \mathbf{reveal}\ \mathsf{h}{\div}2;\ \mathsf{v}{:=}\mathsf{h}{\div}4 \quad\quad \text{``revelation equivalence §9.1''} \\
\sqsubseteq & \mathsf{v}{:=}\mathsf{h}{\div}4\ . \quad\quad \text{``(22), that } \mathbf{reveal}\ \mathsf{h}{\div}2 \sqsubseteq \mathbf{skip}\text{''}
\end{array}
$$

### 9.4 Reveals in sequence

When two or more *HMM*'s are executed sequentially, where the outputs from one are "fed into" another, an observer is able to preserve information from earlier executions to add to information learned by observing later executions. The basic rule expressing the total amount of information leaked is set out next.

**Lemma 9.3** *Sequential reveals*    Let $D_1.\mathsf{v}.\mathsf{h}$ and $D_2.\mathsf{v}.\mathsf{h}$ be distributions over some $\mathcal{X}$ and $\mathcal{Y}$ respectively. Then we have

$$\mathbf{reveal}\ D_1.\mathsf{v}.\mathsf{h}; \mathbf{reveal}\ D_2.\mathsf{v}.\mathsf{h}\quad =\quad \mathbf{reveal}\ (D_1{\times}D_2).\mathsf{v}.\mathsf{h}\ ,$$

where $(D_1{\times}D_2).\mathsf{v}.\mathsf{h}$ is the joint distribution over ordered pairs of $\mathcal{X}\times\mathcal{Y}$, defined as usual so that $(D_1{\times}D_2).\mathsf{v}.\mathsf{h}.(x,y)$ is $D_1.\mathsf{v}.\mathsf{h}.x \times D_2.\mathsf{v}.\mathsf{h}.y$.

*Proof:* This follows directly from the definition of $\mathbf{reveal}\ D.\mathsf{v}.\mathsf{h}$ and sequential composition:

$\[\![\ \mathbf{reveal}\ D_1.\mathsf{v}.\mathsf{h}; \mathbf{reveal}\ D_2.\mathsf{v}.\mathsf{h}\ ]\!]$

$= \quad (\ \sum x,y{:}\ \mathcal{X},\mathcal{Y}\ \bullet\ [\![\{D_1.\mathsf{v}.\mathsf{h}.x\};\{D_2.\mathsf{v}.\mathsf{h}.y\}]\!]\ )\ $ "revelations as summations (6) composition distributes addition"

$= \quad (\ \sum x,y{:}\ \mathcal{X},\mathcal{Y}\ \bullet\ [\![\{D_1.\mathsf{v}.\mathsf{h}.x \times D_2.\mathsf{v}.\mathsf{h}.y\}]\!]\ )\quad\quad\quad$ "(11)"

$= \quad [\![\ \mathbf{reveal}\ (D_1{\times}D_2).\mathsf{v}.\mathsf{h}\ ]\!]\ .\ $ "represent assertion summation as revelation (6)"

□

This rule says that we can simplify two successive reveals into a single reveal where the external values are gathered together and the residual probabilities aggregated as expected, so that overall the result is as though a single *HMM* had been executed, albeit with a modified stochastic matrix. Using this basic rule we can prove the following:

---

[15] With only a selection of laws, sometimes we must omit details in the calculations.

**Lemma 9.4** *Simple sequential rules*

$$\textbf{reveal } D_1.\textsf{v.h}; \textbf{reveal } D_2.\textsf{v.h} \quad = \quad \textbf{reveal } D_2.\textsf{v.h}; \textbf{reveal } D_1.\textsf{v.h} \quad (27)$$

$$\textbf{reveal } E_1.\textsf{v.h}; \textbf{reveal } E_2.\textsf{v.h} \quad = \quad \textbf{reveal } (E_1.\textsf{v.h}, E_2.\textsf{v.h}) \quad (28)$$

$$\textbf{reveal } \textsf{h}; \textbf{reveal } D.\textsf{v.h} \quad = \quad \textbf{reveal } \textsf{h} \quad (29)$$

*Proof:* Using Lem. 9.3, equation (27) follows from the underlying commutativity, and (28) follows from the fact that **reveal** $E.\textsf{v.h}$ equals **reveal** $\{\!\{E.\textsf{v.h}\}\!\}$. For (29) we have from (22) and (33) that **reveal** $\textsf{h}$; **reveal** $D.\textsf{v.h} \sqsubseteq$ **reveal** $\textsf{h}$. For refinement in the other direction we reason

$$\begin{array}{rll}
& \textbf{reveal } \textsf{h} & \\
= & \textbf{reveal } (\textsf{h}, \textsf{h}) & \text{``(26)''} \\
= & \textbf{reveal } \textsf{h}; \textbf{reveal } \textsf{h} & \text{``(28)''} \\
\sqsubseteq & \textbf{reveal } \textsf{h}; \textbf{reveal } D.\textsf{v.h} \ . & \text{``(22)''}
\end{array}$$

□

The rules in Lem. 9.4 formalise our intuition about successive reveals. For example (27) says that that information can be revealed in any order, that revealing two different expressions in succession is the same as revealing a pair containing both expressions (28), and that once $\textsf{h}$ has been revealed entirely then there is nothing more to reveal (29).

The following lemma lists further properties explaining how assertions and revelations interact via sequential composition.

**Lemma 9.5** *Assertions and revelations in sequence*

$$\{p.\textsf{v.h}\}; \textbf{reveal } D.\textsf{v.h} \quad = \quad \textbf{reveal } D.\textsf{v.h}; \{p.\textsf{v.h}\} \quad (30)$$

$$\{G.\textsf{v.h}\} \quad = \quad \{G.\textsf{v.h}\}; \textbf{reveal } G.\textsf{v.h} \quad (31)$$

*Proof:* The first equivalence is shown using a similar proof to that of Lem. 9.3. For (31) we show

$$\begin{array}{rll}
& [\![\{G.\textsf{v.h}\}]\!] & \\
= & [\![\{G.\textsf{v.h}\}]\!] + [\![\{\textsf{F}\}]\!] & \text{``\textbf{abort} is zero of program addition''} \\
= & [\![\{G.\textsf{v.h}\}; \{G.\textsf{v.h}\}]\!] + [\![\{G.\textsf{v.h}\}; \{\neg G.\textsf{v.h}\}]\!] & \text{``separate Boolean assertions (15)''} \\
= & [\![\{G.\textsf{v.h}\}; \textbf{reveal } G.\textsf{v.h}]\!] \ . & \text{``composition over addition; (6)''}
\end{array}$$

□

The first (30) states that revelations and assertions commute, while the second (31) says that after asserting predicate $G.\textsf{v.h}$, no more information can be leaked by revealing the value of $G.\textsf{v.h}$.

## 9.5 Reveals in choice

In a probabilistic choice between two reveal statements, an observer may witness both which revelation was executed as well as the outcome of that statement. We can combine such a choice into a single reveal statement.

**Lemma 9.6** *Choices between reveals*   Let $D_L.\mathsf{v.h}$ and $D_R.\mathsf{v.h}$ be distributions over $\mathcal{X}$ and $p.\mathsf{v.h}$ be a probability; let

$$\mathcal{X}_2 := \ \mathsf{Lft}\ \mathcal{X}\ +\ \mathsf{Rgt}\ \mathcal{X}$$

be the discriminated union of two copies of $\mathcal{X}$, with injection functions therefore of type $\mathsf{Lft}, \mathsf{Rgt}\colon \mathcal{X} \to \mathcal{X}_2$. We have that

$$\begin{array}{c}\mathbf{reveal}\ D_L.\mathsf{v.h} \\ {}_{p.\mathsf{v.h}}\oplus\quad \mathbf{reveal}\ D_R.\mathsf{v.h}\end{array} \quad = \quad \begin{array}{c}\mathbf{reveal}\qquad\mathsf{map.Lft}.(D_L.\mathsf{v.h}) \\ {}_{p.\mathsf{v.h}}\oplus\quad \mathsf{map.Rgt}.(D_R.\mathsf{v.h})\end{array},$$

where the injection-functions' "tagging" of the two distributions has effectively given them disjoint supports.

*Proof:*   Let $D'_L.\mathsf{v.h}, D'_R.\mathsf{v.h}$ be respectively $\mathsf{map.Lft}.(D_L.\mathsf{v.h}), \mathsf{map.Rgt}.(D_R.\mathsf{v.h})$. We have then

$$\begin{aligned}
&\ [\![\ \mathbf{reveal}\ D_L.\mathsf{v.h}\ {}_{p.\mathsf{v.h}}\oplus\ \mathbf{reveal}\ D_R.\mathsf{v.h}\ ]\!] \\
=&\ [\![\ \mathbf{reveal}\ D'_L.\mathsf{v.h}\ {}_{p.\mathsf{v.h}}\oplus\ \mathbf{reveal}\ D'_R.\mathsf{v.h}\ ]\!] &&\text{"Lem. 9.1"} \\[4pt]
=&\ \ [\![\ \{p.\mathsf{v.h}\}; \mathbf{reveal}\ D'_L.\mathsf{v.h}\ ]\!] &&\text{"probabilistic choice"} \\
&+ [\![\ \{1{-}p.\mathsf{v.h}\}; \mathbf{reveal}\ D'_R.\mathsf{v.h}\ ]\!] \\[4pt]
=&\ \sum_{x:\mathcal{X}} &&\text{"revelations are additions of assertions (6);"} \\
&\ [\![\{p.\mathsf{v.h}\}; \{D'_L.\mathsf{v.h}.(\mathsf{Lft}.x)\}]\!] &&\text{additive distributivity} \\
&+ [\![\{1{-}p.\mathsf{v.h}\}; \{D'_R.\mathsf{v.h}.(\mathsf{Rgt}.x)\}]\!] &&\text{of Kleisli composition"} \\[4pt]
=&\ (\ \sum x_2{:}\mathcal{X}_2\ \bullet\ [\![\{(D'_L.\mathsf{v.h}\ {}_{p.\mathsf{v.h}}\oplus D'_R.\mathsf{v.h}).x_2\}]\!]\ ) &&\text{"(11); and } D'_L.\mathsf{v.h}.(\mathsf{Rgt}.x), \\
& &&D'_R.\mathsf{v.h}.(\mathsf{Lft}.x)\ \text{both zero"} \\[4pt]
=&\ [\![\ \mathbf{reveal}\ D'_L.\mathsf{v.h}\ {}_{p.\mathsf{v.h}}\oplus D'_R.\mathsf{v.h}\ ]\!]. &&\text{"addition of assertions as revelation (6)"}
\end{aligned}$$

□

From this lemma we may derive the following laws concerning revelations and probabilistic choice.

**Lemma 9.7** *Simple choice rules*   For probability $p.\mathsf{v.h}$ and distributions $D_1.\mathsf{v.h}$ and $D_2.\mathsf{v.h}$ we have that

$$(\mathbf{reveal}\ D_L.\mathsf{v.h}\ {}_{p.\mathsf{v.h}}\oplus \mathbf{reveal}\ D_R.\mathsf{v.h}) \quad \sqsubseteq \quad \mathbf{reveal}\ (D_L.\mathsf{v.h}\ {}_{p.\mathsf{v.h}}\oplus D_R.\mathsf{v.h})\ .$$

However if the support of $D_L.\mathsf{v.h}$ and $D_R.\mathsf{v.h}$ are disjoint, then the refinement relation is an equality.

*Proof:*   This follows from Lems. 9.6, 9.1.   □

From (21) and Lem. 9.7 we have, for example, that

$$\textbf{skip}\ _{p.\textsf{v}.\textsf{h}}\!\oplus\textbf{skip}\quad =\quad (\textbf{reveal}\ \mathsf{T}\ _{p.\textsf{v}.\textsf{h}}\!\oplus\textbf{reveal}\ \mathsf{F})\quad =\quad \textbf{reveal}\ (\mathsf{T}\ _{p.\textsf{v}.\textsf{h}}\!\oplus\mathsf{F})\ ,$$

thus illustrating the information leakage due to implicit flow.

### 9.6 Composition, probabilistic choice and conditionals

As well as being monotonic in both their program arguments (Thm. 8.1), sequential composition and probabilistic choice –of which conditional choice is a special case– satisfy the following basic laws corresponding to classical probabilistic equalities [28].

**Lemma 9.8** *Basic composition and choice laws* For all programs $S$, $T$ and $R$ and probabilities $p.\textsf{v}.\textsf{h}$ we have the following properites hold.

$$
\begin{align}
S;(T;R) \quad &= \quad (S;T);R & (32) \\
\textbf{skip};S \quad &= \quad S;\textbf{skip}\ =\ S & (33) \\
\textbf{abort};S \quad &= \quad S;\textbf{abort}\ =\ \textbf{abort} & (34) \\
(S\ _{p.\textsf{v}.\textsf{h}}\!\oplus T) \quad &= \quad (T\ _{1-p.\textsf{v}.\textsf{h}}\!\oplus S) & (35) \\
(S\ _{p.\textsf{v}.\textsf{h}}\!\oplus T);R \quad &= \quad (S;R\ _{p.\textsf{v}.\textsf{h}}\!\oplus T;R) & (36) \\
(S\ _1\!\oplus T) \quad &= \quad S & (37)
\end{align}
$$

Additionally, for any $R$ satisfying both $R;\{p.\textsf{v}.\textsf{h}\} = \{p.\textsf{v}.\textsf{h}\};R$ and $R;\{1-p.\textsf{v}.\textsf{h}\} = \{1-p.\textsf{v}.\textsf{h}\};R$ we have that $R$ distributes from the left into a choice with probability $p.\textsf{v}.\textsf{h}$:

$$R;(S\ _{p.\textsf{v}.\textsf{h}}\!\oplus T)\quad =\quad (R;S\ _{p.\textsf{v}.\textsf{h}}\!\oplus R;T)\ . \tag{38}$$

□

Since both assertions (11) and reveals (30) commute over assertions, equation (38) gives us that they distribute to the right over arbitrary probabilistic (and conditional) choices. Additionally, commutativity of constant assertions over all statements (12) means that all programs distribute over choices in which the probability is constant. We can also derive, for example, the following properties:

$$
\begin{align}
S\ _{p.\textsf{v}.\textsf{h}}\!\oplus S \quad &\sqsubseteq \quad S & (39) \\
S\ _{p.\textsf{v}}\!\oplus S \quad &= \quad S & (40) \\
\textbf{if}\ G\ \textbf{then}\ S\ \textbf{else}\ T \quad &= \quad \textbf{reveal}\ G;\textbf{if}\ G\ \textbf{then}\ S\ \textbf{else}\ T & (41)
\end{align}
$$

The first two follow from left distributivity (33,36,21) and Lem. 9.7. The last follows from (38,31).

## 9.7 Rules for general iteration

Recall that we write **while** $p$.v.h **do** $S$ **od** for a general iteration of $S$, with probability $p$.v.h of exiting the loop on each iteration. From Thm. 8.1 we have that such loops are monotonic on their program argument. Additionally, from its least-fixed-point semantics least fixed point we have

**Lemma 9.9** *Fixed point rule* If $(S;W)\ _{p.\textsf{v.h}}\oplus \textbf{skip} = W$ then we have the refinement (**while** $p$.v.h **do** $S$ **od**) $\sqsubseteq W$.
*Proof:* From the Tarski fixed-point theorem [46] wrt the order ($\leq$), and that the loop is a least fixed point, we have immediately

$$(S;W)\ _{p.\textsf{v.h}}\oplus \textbf{skip}\ \leq\ W \quad \text{implies} \quad (\textbf{while}\ p.\textsf{v.h}\ \textbf{do}\ S\ \textbf{od})\ \leq\ W\ .\quad (42)$$

The result then follows immediately from the two inclusions $(=) \subseteq (\leq) \subseteq (\sqsubseteq)$.
□

In a specification task, however, the goal is typically to implement a specification by an iteration, i.e. to establish a refinement in the opposite direction. For terminating iterations we have this rule:

**Corollary 9.1** *Termination iteration* If **while** $p$.v.h **do** $S$ **od** terminates with probability one, and $(S;W)\ _{p.\textsf{v.h}}\oplus\textbf{skip} = W$, then **while** $p$.v.h **do** $S$ **od** $= W$.
*Proof:* We adapt the proof of Lem. 9.9, noting that if the loop terminates it is ($\leq$)-maximal and hence the *rhs* ($\leq$) in (42) must in fact be an equality. □

Termination is usually shown by exhibiting a probabilistic variant over the state [19, 34, 28]; a straightforward simple case is when the loop's exit probability is bounded away from zero, in particular **while** $k$ **do** $\cdots$ for any constant $k<1$.

## 9.8 Small example: one guess at a password

We have a hidden password p chosen from three possibilities $\mathcal{P}:=\{p_1,p_2,p_2\}$. This fragment describes an attacker's single guess, uniformly chosen:

$$\|[\ \textbf{vis}\ \textsf{g};\ \textsf{g}{:}{\in}\{\!\!\{p_1,p_2,p_3\}\!\!\};\ \textbf{reveal}\ \textsf{g}{=}\textsf{p}\ ]\|\ .$$

Local visible value g is chosen from the uniform distribution $\{\!\!\{p_1,p_2,p_3\}\!\!\}$, and then it is used as a guess. Note that if the guess is correct, then T is revealed which –in itself– does not reveal the password's value: that latter is *then* learned by deduction, from the program's code and the fact that g is visible. If g had been hidden, we would know only that the guess had succeeded, but still not the value of p.

We now show how algebra can be used to convert "operational" descriptions like the above into less obvious but more calculationally convenient forms, in this case a single **reveal** statement; and in §10 we will see how useful this equivalence turns out to be. For now, we reason

$$\||[ \text{ vis g; g:} \in \{\!\{p_1, p_2, p_3\}\!\}; \text{ reveal g=p } ]\|\|$$

| | | |
|---|---|---|
| $=$ | $\||[$ **vis** g;<br>  g:=$p_1 \oplus$ g:=$p_2 \oplus$ g:=$p_3$;<br>  **reveal** g=p<br>$]\|\|$ | "split visible choice using (7);<br>note choices ($\oplus$) are uniform<br>by convention, i.e. $(_{1/3}\oplus)$<br>in this case" |
| $=$ | $\||[$ **vis** g;<br>  g:=$p_1$; **reveal** g=p;<br>$\oplus$ g:=$p_2$; **reveal** g=p;<br>$\oplus$ g:=$p_3$; **reveal** g=p<br>$]\|\|$ | "left distributivity (36)" |
| $=$ | $\||[$ **vis** g;<br>  g:=$p_1$; **reveal** $p_1$=p;<br>$\oplus$ g:=$p_2$; **reveal** $p_2$=p;<br>$\oplus$ g:=$p_3$; **reveal** $p_3$=p<br>$]\|\|$ | "replace expressions by those of equal value (§9.1);<br>e.g. in the first branch g:=$p_1$ establishes<br>that $p_1$=g, so that g can be<br>replaced by $p_1$ in the **reveal**" |
| $=$ | $\||[$ **vis** g; g:=$p_1$ $]\|\|$; **reveal** $p_1$=p<br>$\oplus$ $\||[$ **vis** g; g:=$p_2$ $]\|\|$; **reveal** $p_2$=p<br>$\oplus$ $\||[$ **vis** g; g:=$p_3$ $]\|\|$; **reveal** $p_3$=p | "shift scope (§9.1), since g is<br>no longer free in the **reveal**'s" |
| $=$ | **reveal** $p_1$=p $\oplus$ **reveal** $p_2$=p $\oplus$ **reveal** $p_3$=p | "(9), (8) and (33)" |
| $=$ | **reveal** $(p_1, p_1$=p$)$ $\oplus$ **reveal** $(p_2, p_2$=p$)$ $\oplus$ **reveal** $(p_3, p_3$=p$)$ | "Lem. 9.1" |
| $=$ | **reveal** $(p_1, p_1$=p$) \oplus (p_2, p_2$=p$) \oplus (p_3, p_3$=p$)$ , | "Lem. 9.7" |

giving a single **reveal** whose expression-part we manipulate further, at (46) below. Note that it is the appeal to (7) that relies on g's being visible: if it were not, then the implicit flow introduced by the first step would represent a leak, invalidating the equality.

## 10 Extended example: iterative reasoning

We now demonstrate our treatment of iteration, reusing the simple password-guessing attack within a loop.

### 10.1 A password attack: specification

We assume a set of passwords $\mathcal{P}$ and a hidden variable p:$\mathcal{P}$ containing the (current) password; let $\mathbb{P}_N\mathcal{P}$ be the set of all size-$N$ subsets of $\mathcal{P}$. A typical attack would be to choose one of those sets of potential passwords, and then to try them all in a "bulk attack" as in the program fragment

$$\||[ \text{ vis G; G:} \in \lfloor \mathbb{P}_N\mathcal{P} \rfloor; \text{ reveal } \{\mathsf{p}\} \cap \mathsf{G} \ ]\|\| \ . \tag{43}$$

(We omit the typing G: $\mathbb{PP}$, to reduce clutter.) The statement G:$\in \lfloor \mathbb{P}_N \mathcal{P} \rfloor$ makes a uniform choice of size-$N$ subset of $\mathcal{P}$, assigning it to G. We are assuming that $N$ is strictly less than the size $P$ of $\mathcal{P}$.

The **reveal** {p}∩G reveals either {p}, if the attack succeeds, or the empty set ∅ if it does not. That is, the outcome of fragment (43) above is either to say "the hidden password is p" (a successful attack, revealing {p}) or "the hidden password is not in G" (an unsuccessful attack, revealing ∅) since, in the latter case we do know the visible attack-set G even though the attack failed. As a specification, it abstracts from precisely how the passwords are tried, in what order, or whether possibly repeated: it says only thay they *are* tried.

Now suppose the incoming distribution of p is some $\pi$: $\mathbb{D}\mathcal{P}$; then the program fragment above produces an output hyper Π: $\mathbb{D}^2\mathcal{P}$ comprising a distribution of distributions over $\mathcal{P}$. (Note that the output hyper contains no G component, because G is local.) If we calculated this with our semantics (although we omit the calculations here), we would find two kinds of inners in its support, namely

**success** A $p$-indexed family of point inner distributions $\{\!|p|\!\}$ each itself with outer probability $N(\pi.p)/P$, the probability $\pi.p$ that $p$ was the password, but multiplied by the probability $N/P$ that it was in the uniformly chosen attack-set G of size $N$.

**failure** A $G$-indexed set of inner distributions of support-size $P{-}N$, each such distribution derived by conditioning $\pi$ on *not* being in the set $G$ and having outer probability $(1{-}\pi.G)/\mathsf{C}_N^P$, the probability that this particular $G$ was chosen for the attack-set multiplied by the probability that the password was not in it.

As a check, we note that the outer probabilities sum to one, as they should since the specification program is terminating: we have

$$\begin{array}{rl} & \sum_p N(\pi.p)/P + \sum_G (1{-}\pi.G)/\mathsf{C}_N^P \\ & \sum_p N(\pi.p)/P + \sum_G 1/\mathsf{C}_N^P - \sum_G \pi.G/\mathsf{C}_N^P \\ = & N/P + \mathsf{C}_N^P/\mathsf{C}_N^P - \sum_p \mathsf{C}_{N-1}^{P-1}(\pi.p)/\mathsf{C}_N^P \\ = & 1 \ . \end{array}$$

Finally, if for example we assume that the incoming distribution $\pi$ is uniform over $\mathcal{P}$, then the Bayes Risk before the attack is $1 - 1/P$ and, after the attack, it has been reduced to the conditional Risk $P{\times}N(\pi.p)/P{\times}0 + \mathsf{C}_N^P{\times}((1{-}N/P)/\mathsf{C}_N^P){\times}(1 - 1/(P{-}N))$, that is reduced to $1 - (N{+}1)/P$.

## 10.2 A password attack: implementation

We suppose a simple-minded actual attacker who chooses single passwords uniformly at random, possibly with repetition and, after each attack, has some fixed probability $c$ of giving up. This would be described by the fragment

$$\begin{array}{l} \textbf{while } c \textbf{ do} \\ \quad |\![ \textbf{ vis } g; \ g{:}{\in}\mathcal{P}; \ \textbf{reveal } g{=}p \ ]\!| \\ \textbf{od} \ . \end{array} \qquad (44)$$

A complete analysis of (44) is combinatorially complex, having an output hyper comprising inner distributions over subsets of $\mathcal{P}$ of all possible sizes and –as such– would be difficult to reason about within a larger system. More practical would be to determine, once and for all, whether (44) is an implementation of (43), that is is at least as secure as (43) and then ever after to use the simpler (43) in larger analyses. Since (44) is parametrised by $c$, we might in fact ask

What is the largest value of probability $c$ for which (43) $\sqsubseteq$ (44)?

## 10.3 Example refinement analysis: the simplest case

To illustrate the approach, we address the above question in the very simple case where $\mathcal{P}=\{p_1, p_2, p_3\}$ is of size 3, and our specification describes a "bulk attack" of size $N=1$.[16] Thus we are asking for the largest $c$ that achieves the refinement

$$\|[\text{ vis g};\ \text{g}{:}\in\mathcal{P};\ \textbf{reveal}\ \text{g=p }]\| \quad \sqsubseteq \quad \begin{array}{l}\textbf{while } c \textbf{ do} \\ \quad \|[\text{ vis g};\ \text{g}{:}\in\mathcal{P};\ \textbf{reveal}\ \text{g=p }]\| \\ \textbf{od}\ . \end{array} \quad (45)$$

We do this in two stages: the first is to hypothesise a parametrised straight-line equivalent for the loop, then synthesising a condition on the parameters that makes it satisfy the fixed-point equation of Cor. 9.1.

As in Lem. 9.6, we introduce a discriminated union $\mathcal{P}^? := \text{is } \mathcal{P} + \text{isn't } \mathcal{P} + \text{nix}$ which, used in **reveal** commands, will allow us to reveal what p is, what it is not, and –for algebraic convenience– to reveal nothing at all.

In our simple case here of $\mathcal{P}$ having just three elements, therefore $\mathcal{P}^?$ has seven. Further exploiting $\mathcal{P}$'s size of three, for any $p$ in $\mathcal{P}$ we write $p_+$ for one of the values $p$ is not, and $p_-$ for the other. With this approach we can express *lhs* (45) without its local block and the guess variable g: for that, we return to our example calculation of §9.8 giving **reveal** $(p_1, p_1{=}\text{p}) \oplus (p_2, p_2{=}\text{p}) \oplus (p_3, p_3{=}\text{p})$. We can recode this directly using Lem. 9.1: it becomes just

$$\textbf{reveal } \{\!\!\{\text{is p, isn't p}_+, \text{isn't p}_-\}\!\!\} \ . \qquad (46)$$

We return to the synthesis of the loop's straight-line equivalent, supposing it has the form

$$\textbf{reveal } \{\!\!\{\text{is p}^{@x}, \text{isn't p}_+{}^{@\frac{y}{2}}, \text{isn't p}_-{}^{@\frac{y}{2}}, \text{nix}^{@z}\}\!\!\} \qquad (47)$$

for some probabilities $x+y+z=1$ that we have to determine. This reveals what p is with probability $x$, what p is not with probability $y/2+y/2 = y$; and with probability $z$ it reveals nothing at all.

---

[16]In this simple case a bulk attack of size $N=2$ is uninteresting, because it would reveal everything: either what the password is (if p∈G) or two values that it is not (if p∉G). In the latter case we would deduce p's value anyway, by elimination.

Our synthesising equality is then given by Cor. 9.1, because the loop with its constant $c$ terminates; that is we require

$$\textbf{reveal } \{\!\!\{\text{is p}^{@x}, \text{isn't p}_+^{@\frac{y}{2}}, \text{isn't p}_-^{@\frac{y}{2}}, \text{nix}^{@z}\}\!\!\} \quad \} \leftarrow (47)$$

$$= \quad \begin{array}{l} \textbf{reveal } \{\!\!\{\text{is p}, \text{isn't p}_+, \text{isn't p}_-\}\!\!\}; \\ \textbf{reveal } \{\!\!\{\text{is p}^{@x}, \text{isn't p}_+^{@\frac{y}{2}}, \text{isn't p}_-^{@\frac{y}{2}}, \text{nix}^{@z}\}\!\!\} \\ _c\oplus \textbf{ skip }, \end{array} \quad \begin{array}{l} \longleftarrow \text{ loop body} \\ \} \leftarrow (47) \\ \longleftarrow \text{ loop exit} \end{array}$$

whose right-hand side we can simplify with the revelation laws from §9, in particular Lems. 9.1,9.3&9.7. That gives

$$\textbf{reveal} \quad \{\!\!\{ \begin{array}{lll} \text{is p} & @ \ c(x + 2y/3 + z/3) & , \\ \text{isn't p}_+ & @ \ c(y/6 + z/3) & , \\ \text{isn't p}_- & @ \ c(y/6 + z/3) & , \\ \text{nix} & @ \ 1{-}c & \}\!\!\} \end{array} ,$$

and that should be equal to the left-hand side, the original (47). Since $z = 1{-}c$ trivially, we concentrate $\text{p}_+$ case to obtain $y/2 = c(y/6 + 2(1{-}c)/3)$, so that $y = 2c(1{-}c)/(3{-}c)$, whence $x = c{-}y = c(1{+}c)/(3{-}c)$.[17]

## 10.4 Establishing the $c$-optimal refinement: the second stage

We now want to find the largest value of $c$ that allows

$$\|[ \textbf{ vis g}; \text{ g:}{\in}\mathcal{P}; \textbf{ reveal g=p }]\|$$
$$\sqsubseteq \quad \textbf{reveal } \{\!\!\{\text{is p}^{@x}, \text{isn't p}_+^{@\frac{y}{2}}, \text{isn't p}_-^{@\frac{y}{2}}, \text{nix}^{@z}\}\!\!\}$$

where $x, y, z$ have the $c$-determined values calculated above: we recall the remark above at (46) about formulating our specification as a simple revelation, without needing a local variable g. That gives the equivalent goal

$$\textbf{reveal } \{\!\!\{\text{is p}^{@\frac{1}{3}}, \text{isn't p}_+^{@\frac{1}{3}}, \text{isn't p}_-^{@\frac{1}{3}}\}\!\!\}$$
$$\sqsubseteq \quad \textbf{reveal } \{\!\!\{\text{is p}^{@x}, \text{isn't p}_+^{@\frac{y}{2}}, \text{isn't p}_-^{@\frac{y}{2}}, \text{nix}^{@z}\}\!\!\}$$

---

[17]Working through this and extracting the arithmetic results in the following table:

↓ effective joint revelation

| | | |
|---|---|---|
| is $p$, is $p$ | with prob. $x/3$ equivalent to revealing just | is $p$ |
| is $p$, isn't $p_+$ | with prob. $y/6$ equivalent to revealing just | is $p$ |
| is $p$, isn't $p_-$ | with prob. $y/6$ equivalent to revealing just | is $p$ |
| is $p$, nix | with prob. $z/3$ equivalent to revealing just | is $p$ |
| isn't $p_+$, is $p$ | with prob. $x/3$ equivalent to revealing just | is $p$ |
| isn't $p_+$, isn't $p_+$ | with prob. $y/6$ equivalent to revealing just | isn't $p_+$ |
| isn't $p_+$, isn't $p_-$ | with prob. $y/6$ equivalent to revealing just | is $p$ |
| isn't $p_+$, nix | with prob. $z/3$ equivalent to revealing just | isn't $p_+$ |
| isn't $p_-$, is $p$ | with prob. $x/3$ equivalent to revealing just | is $p$ |
| isn't $p_-$, isn't $p_+$ | with prob. $y/6$ equivalent to revealing just | is $p$ |
| isn't $p_-$, isn't $p_-$ | with prob. $y/6$ equivalent to revealing just | isn't $p_-$ |
| isn't $p_-$, nix | with prob. $z/3$ equivalent to revealing just | isn't $p_-$ . |

For this we refer to Lem. 9.1, whose $D$ is effectively the *lhs* above: written as a matrix, it would be

|      | is $p_1$ | isn't $p_1$ | is $p_2$ | isn't $p_2$ | is $p_3$ | isn't $p_3$ | nix |
|------|----------|-------------|----------|-------------|----------|-------------|-----|
| $p_1$ | 1/3 | 0   | 0   | 1/3 | 0   | 1/3 | 0 |
| $p_2$ | 0   | 1/3 | 1/3 | 0   | 0   | 1/3 | 0 |
| $p_3$ | 0   | 1/3 | 0   | 1/3 | 1/3 | 0   | 0 |

(48)

We need a 7×7 stochastic matrix $F$, that is a function $\mathcal{P}^? \to \mathbb{D}\mathcal{P}^?$ which, when multiplied after $D$, gives the *rhs* above, that is

|      | is $p_1$ | isn't $p_1$ | is $p_2$ | isn't $p_2$ | is $p_3$ | isn't $p_3$ | nix |
|------|----------|-------------|----------|-------------|----------|-------------|-----|
| $p_1$ | $x$ | 0     | 0   | $y/2$ | 0   | $y/2$ | $z$ |
| $p_2$ | 0   | $y/2$ | $x$ | 0     | 0   | $y/2$ | $z$ |
| $p_3$ | 0   | $y/2$ | 0   | $y/2$ | $x$ | 0     | $z$ |

.

The columns of the latter must be interpolations of columns of the former, thus the first *rhs* column $[x,0,0]$ cannot contain non-zero contributions from any other than the first *lhs* column $[1/3,0,0]$.[18] Hence $x \leq 1/3$ and, since we are trying to maximise $c$ we maximise $x$ also by setting $x := 1/3$.[19] Similar reasoning then establishes that the second *rhs* column $[0, y/2, y/2]$ must be obtained by taking proportion $3y/2$ of the second *lhs* column; and then the last *rhs* column is made by combining proportions $1 - 3y/2$ of each of columns 1,3,5 on the *lhs*.

Since $x=1/3$ entails $c(1+c)/(3-c)=1/3$, that is $c \approx 0.53$, we have established our desired (45) with $c$ taking that value (or less), independently of the distribution with which the hidden p might have been chosen.

## 11 Related work

### 11.1 *HMM*'s, algebra and noninterference

*Hidden Markov Models* [22] have a long history and many practical applications; their conceptual connection to noninterference suggests that their algorithmic methods might be of use here. That is, extant *HMM* techniques could be used for efficient numerical calculation of whether some $\boldsymbol{T}_S, \boldsymbol{E}_S$, a specification, was secure enough for our purposes: once that was done, the refinement relation established via program-algebra could ensure that an implementation $\boldsymbol{T}_I, \boldsymbol{E}_I$ was at least as secure as that *without* requiring a second numerical calculation. The advantage of this is that the first calculation, over a smaller and more abstract system, is likely to be much simpler than the second would have been.

There are techniques based on the manipulation of "graphical models" to represent Bayesian networks in alternate equivalent ways: these are similar in

---

The probabilities in the text come from adding the final column in groups.

[18] We write transposed columns horizontally as rows between brackets [·] instead of parentheses.

[19] The function $x := c(1+c)/(3-c)$ is monotonic for $0 \leq c \leq 1$.

spirit to our algebraic manipulations [7], although there the motivation is usually to find more efficient algorithms.

The application of *HMM*'s to noninterference security is recent: originally, noninterference was qualitative [16]. Probabilistic noninterference [44, 11] is a generalisation of that idea to provide weaker statements concerning an attacker's ability to guess high security state by observing the behaviour and pattern of observables. Variations of the idea have been studied extensively for concurrent systems [43, 48] and taking computational issues into account [6].

The definition of our space $\mathbb{D}(\mathcal{V} \times \mathbb{D}\mathcal{H})$ and its refinement order draws inspiration from constructions and techniques already present in the literature. The monad is Giry/Kanotorovich [15, 47], and the refinement order is related to the theory of inhomogeneous Markov Chains [12].

## 11.2 Compositionality, information theory and assorted entropies

A compelling approach to quantitative security is to use information-theoretic measures to compare the (e.g. Shannon) entropy of the hidden variables' *a priori* distribution (e.g. their incoming values) and their *a posteriori* distribution once the program has executed [10, 11, 3]; recently this has been applied to iterating programs as well [26, 40]. But *compositionality* is crucial: given that one program is more secure than another according to some entropy-based criterion, how do we know that inequality is preserved in a larger context?

We have shown earlier [27] refinement has two key properties for compositional entropy-based reasoning: it is preserved by contexts; and it implies nondecrease for an assortment of entropies, including Shannon Entropy, Guessing Entropy, Bayes Risk and Marginal Guesswork. Perhaps it applies to others [9].

Thus our work here is part of a larger program to unite earlier work in quantitative information flow (or escape) [3, 24] in channels, as models of computation, with a *denotational* presentation of program semantics based on *HMM*'s including a *compositional* refinement relation that compares these quantitative measures between programs, specifically between specifications and their purported implementations. By considering iterations, we are extending our own earlier work [27] in a way that relates to others' work on quantitative information flow from iterations [26, 40] much as in the way described above.

Compositionality "within" a program addresses the question of whether security established for a component is preserved when embedded in a larger context [8]. Compositionality "between" programs, as we do here, addresses the question of whether two programs' relative security is preserved when they are both placed in the same context: this latter is less common.

A representative example of others' doing so is recent work by Yasuoka and Terauchi [51] in which computational hardness is analysed. They consider deterministic sequential straight-line programs. i.e. without probabilistic or demonic choice and without loops, but that nevertheless operate in a quantitative context (i.e. having input *distributions* rather than simply input values). Since the programs have no probabilistic choices, those authors are able to reduce the (analogue of) the secure-refinement relation to the qualitative noninterference comparison of the programs. This is a special case of our general conjecture concerning the promotion of qualitative results to quantitative results provided demonic choice is replaced by uniform choice [27, Sec.8.1]: if there is no demonic choice, there is nothing to replace and so the program is unchanged.

These authors look for a relation guaranteeing the correct entropy ordering (for all incoming distributions) wrt a selection of entropies, as we do, and they address the computational hardness of validating that relationship in particular cases. We address with compositional *closure* the additional question of how weak such a relation can be [27].

## 12 Summary, conclusions, prospects

Earlier we built the core of a programming algebra for probability and noninterference: here we have extended it to include iteration and nontermination; and we solved the technical problem of incompleteness, that arose in the process, by introducing a simpler "termination" order that allowed us to remain with discrete distributions. Further, we have shown how the semantics is related to *HMM*'s, an existing consensus of how such application domain should be handled and analysed.

The formalist rigour of program semantics, however, can make unusual demands on traditional mathematical presentations: a programming language is interpreted inductively in a structured space equipped with operators corresponding to the constructors of that language. In particular, sequential programs with any kind of nondeterminism (whether demonic, probabilistic or some other) are often interpreted as functions of type $SS{\rightarrow}\mathbb{K}SS$ where $SS$ is the state space and $\mathbb{K}$ is some type constructor (or functor) expressing the non-determinism. Thus our first contribution in detail was to (re-)interpret *HMM*'s in this style (in §3), where $SS$ was $\mathcal{V}{\times}\mathbb{D}\mathcal{H}$ and $\mathbb{K}$ became $\underline{\mathbb{D}}$ (in §3.4). We made some small programming-motivated extensions to the *HMM* model, in particular adding *visible* variables to the state so that the most recent observation is carried forward into the next operation. The second extension was allowing *iterations* and hence, potentially, computations that might not terminate (thus $\underline{\mathbb{D}}$ rather than $\mathbb{D}$).

In constructing the semantic operations we built-in *perfect recall* and *implicit flow*, which are security assumptions about the power of the attacker. This can be controversial: in general one can choose to impose these or not. We *did* impose them because we have argued extensively elsewhere [36, 37] that a compositional definition of program refinement is not possible otherwise.[20] Perfect recall in particular, however, does seem a good fit for *HMM*'s independently of the the refinement argument, since the knowledge gained from observations, once emitted from the output-side of an *HMM*, cannot be expunged from the attacker's repertoire by any kind of overwriting subsequently.

---

[20]We did not have space to repeat those arguments here.

Our second contribution was to work-around ($\sqsubseteq$)-incompleteness by using an alternative, more specialised order ($\leq$), showing that a program algebra including iteration is feasible (§9); and our third contribution was to argue by example that the resulting source-level reasoning is promising (§10).

There are two immediate prospects for further work. One that in practice we would like to answer questions like the one posed in §10.2 for general guesses of size $N$ and large password spaces $\mathcal{P}$, and many other similar. For this we would need tool support both for the semantics (i.e. given a program, determine its meaning) and for establishing refinement (i.e. whether this meaning refined by that one) in a probabilistic setting [23, 30].

The other prospect is to complete our semantic space to proper measures, in fact to follow the approach outlined in §6.2. Beyond compositionality of ($\sqsubseteq$) we want its compositional *closure*, already achieved for straight-line programs, guaranteeing that the refinement relation is not unnecessarily strong; but that argument required (analytical) closure/compactness of a set of finite, discrete probability distributions in a metric space [27]; and to do that here, with the extra feature of iterations that generate chains of approximants, seems to make the move to measures inevitable.

Finally, our longer-term aim is to add demonic choice to the model for e.g. demonic scheduling that takes into account what the adversary can, and cannot see [3]. We have done this for qualitative systems [36, 37] and we have earlier combined demonic- and probabilistic choice *without* hiding [20, 39, 28]. The technique of convex closure, useful for that, generates uncountably many interpolated distributions: it is a second reason we are likely to need measures, and so we hope to exploit the structures developed for this paper at that later point.

---

These appendices contain preliminary, background material intended to support further extension of the main results, above, in a separate, subsequent publication; they were written after the main report and are not strictly part of it.

---

# A  The Kantorovich metric and its related probability monad

We begin by recalling the notation and structures we have been using for the discrete case. Let $\boldsymbol{D}$ be a finite set, of size some $N$; then $\mathbb{D}\boldsymbol{D}$ is the set of discrete distributions over it.[21] The *Manhattan metric* between two of those distributions $\delta_{\{1,2\}}$ is then given by $\boldsymbol{m}.\delta_1.\delta_2 := \sum_d |\delta_1.d - \delta_2.d|$ for $d{:}\,\boldsymbol{D}$. Scaled by $1/N$, the metric becomes 1-bounded; in any case, the usual Euclidean topology is induced, on $\mathbb{R}^N$ effectively, and the space is compact.

Now more generally, given a metric space its *Borel algebra* is the smallest sigma-algebra containing the open sets; and the combination of such a space and the Borel algebra is a *measurable space*. The space $(\mathbb{D}\boldsymbol{D}, \boldsymbol{m})$ of discrete distributions, above, is such a metric space; and we can therefore define *measure spaces* $(\mathbb{D}\boldsymbol{D}, \mathcal{B}(\mathbb{D}\boldsymbol{D}, \boldsymbol{m}), \mu)$ over $\mathbb{D}\boldsymbol{D}$, where $\mathcal{B}$ has constructed the required Borel algebra: these $\mu$ are distributions of distributions over $\boldsymbol{D}$, what we have been calling *hypers*.[22] The special case of *discrete* hypers could be said to be the set $\mathbb{D}^2\boldsymbol{D}$.

Because $(\mathbb{D}\boldsymbol{D}, \boldsymbol{m})$ is compact and its metric 1-bounded (once scaled), the *Kantorovich metric* construction can be used to "lift" the underlying metric $\boldsymbol{m}$ to a new metric on the measures $\mu$ themselves, i.e. giving a distance between any $\mu_{\{1,2\}}$ [13] — and that lifted metric is again 1-bounded and makes the space of measures compact. This gives a new 1-bounded and compact metric space of measures over distributions over $\boldsymbol{D}$, which we write $\mathbb{M}\mathbb{D}\boldsymbol{D}$: it depends implicitly on the underlying metric $\boldsymbol{m}$ on $\mathbb{D}\boldsymbol{D}$. And because 1-boundedness and compactness is re-established, the process can be repeated, going on to form e.g. $\mathbb{M}^2\mathbb{D}\boldsymbol{D}$ etc.

In fact if $\boldsymbol{D}$ is itself given the discrete metric $\boldsymbol{d}.d_1.d_2 := 1$, then the Kantorovich construction gives (up to scaling) the Manhattan metric on $\mathbb{D}\boldsymbol{D}$ that we have already chosen, and so for $\mathbb{M}\mathbb{D}\boldsymbol{D}$ we could just as well write $\mathbb{M}^2\boldsymbol{D}$; but as a mnemonic aid we continue to use $\mathbb{D}$ where it applies.

---

[21] We use bold $\boldsymbol{D}$ for the underlying set, rather than e.g. a calligraphic $\mathcal{X}$ as in the main report, for notational reasons explained below.

[22] Note that $\delta_{\{1,2\}}$ were measures, actually discrete distributions over $\boldsymbol{D}$, whereas $\mu$ is a measure over $\mathbb{D}\boldsymbol{D}$.

## A.1 Notation and conventions for measures and the Kantorovich functor

In the presentation further below we will be using many (at least six) different but related measurable spaces, based on metric spaces as above, with each one introducing at least seven derived variables: the space itself; the underlying set; the metric; the induced Borel-algebra; the measurable sets in that algebra; a measure (or measures) defined on the algebra; and finally a variable ranging over the underlying set itself. To achieve some (local) naming consistency, we follow these conventions systematically:

- The underlying set will be named in bold upper-case Roman, thus $\boldsymbol{A}$, and its elements will be lower-case Roman so that for example $a \in \boldsymbol{A}$.

- The associated metric will be in bold lower-case Roman, thus $\boldsymbol{a} \in \boldsymbol{A}^2 \to \mathbb{R}$.

- The metric space as a whole will be in underlined upper-case Roman, thus $\underline{A} = (\boldsymbol{A}, \boldsymbol{a})$.

- The Borel-algebra induced by the metric $\boldsymbol{a}$ on $\boldsymbol{A}$ will be the in corresponding Roman calligraphic, thus $\mathcal{A} \subseteq \mathbb{P}\boldsymbol{A}$.

- The measurable sets in $\mathcal{A}$ will be in upper-case Roman, so that $a \in A \in \mathcal{A}$.

---

- Thus the measurable space derived from $(\boldsymbol{A}, \boldsymbol{a})$ will be $(\boldsymbol{A}, \mathcal{A})$.

---

- Measures over $(\boldsymbol{A}, \mathcal{A})$ will be in lower-case Greek, so that $\alpha.A \in \mathbb{R}$ and then $(\boldsymbol{A}, \mathcal{A}, \alpha)$ is the measurable space written in full. Abusing the types, we sometimes write $\underline{A}$ for the set of those measures, thus by $\alpha \in \underline{A}$ meaning that $(\boldsymbol{A}, \mathcal{A}, \alpha)$ is a measurable space.

- The Kantorovich construction taking metric space $\underline{A}$ to the metric space of measures over it will be written $\mathbb{M}\underline{A}$. By a similar abuse we write $\beta \in \mathbb{M}\underline{A}$ to mean that $\beta$ is one of those measures.

- Given a function $f \colon \boldsymbol{A} \to \boldsymbol{B}$ we write $\mathbb{M}f$ for the corresponding function between (the measures in) $\mathbb{M}\underline{A}$ and $\mathbb{M}\underline{B}$ with the usual definition
$\mathbb{M}f.\alpha.B := \alpha.(f^{-1}.B)$.

We will base our calculations below on van Breugel's presentation of the Kantorovich monad, whose relevant results we now summarise (but in the notation we have established above); the page references are to van Breugel's publication [47]. In fact we will be using *subprobability* measures to take nontermination into account [47, Sec.5.3], and will assume that throughout the following.

**Lemma A.1** *Facts concerning Kantorovich subprobability-measure monads and metrics*

**p13** The unit and multiplication of the monad on $\mathbb{M}$ are defined as for the Giry monad, with an argument necessary that the Giry-style definition of multiplication is meaningful for Kantorovich [50]. The unit $\boldsymbol{\eta}$ is the "make a point measure" function pnt and the multiplication $\boldsymbol{\mu}$ is the "average" function avg.

**p13** If metric space $\underline{X}$ is compact, then $\mathbb{M}\underline{X}$ is compact as well.

**p13** If metric space $\underline{X}$ is complete, then $\mathbb{M}\underline{X}$ is complete as well if we restrict ourselves to tight measures.

**p13** If a function $f$ is nonexpansive, then so is $\mathbb{M}f$.

**p13** Both the unit $\boldsymbol{\eta}$ and the multiplication $\boldsymbol{\mu}$ of the $\mathbb{M}$-monad are nonexpansive.

**p14** $\langle \mathbb{M}, \boldsymbol{\eta}, \boldsymbol{\mu} \rangle$ is a monad on the category of 1-bounded compact metric spaces with nonexpansive functions between them.

$\square$

# B Antisymmetry of refinement

## B.1 Refinement of measures

We begin by revisiting our definition of refinement, placing it in the measure context.

**Definition B.1** *Entropy refinement of measure-hypers* Refinement is a relation on measures $\Delta$ in $\mathbb{MD}\boldsymbol{X}$, i.e. hypers, that –informally– merges elements of $\mathbb{D}\boldsymbol{X}$ together based on the weights assigned to them by $\Delta$. [23]

More precisely we have the following:

1. Start with a finite set $\boldsymbol{X}$ with the discrete metric (such as our program state-space).

2. From (1) construct $\mathbb{D}\boldsymbol{X} = \mathbb{M}\boldsymbol{X}$ with the Kantorovich/Manhattan metric. These are our discrete distributions.

3. From (2) construct $\mathbb{MD}\boldsymbol{X} = \mathbb{M}^2\boldsymbol{X}$ with the Kantorovich metric. These are our hypers.

4. From (3) construct $\mathbb{M}^2\mathbb{D}\boldsymbol{X} = \mathbb{M}^3\boldsymbol{X}$ with the Kantorovich metric. These are our supers.

5. From (4) construct $\mathbb{M}^3\mathbb{D}\boldsymbol{X} = \mathbb{M}^4\boldsymbol{X}$ with the Kantorovich metric, which is where the "mega" lives that is used in the conjecture supporting transitivity.

---

[23] We retain the use of upper-case Greek letters for hypers, in spite of the conventions above, for consistency with the main paper.

6. For two hypers $\Delta_{\{S,I\}}$ say that $\Delta_S$ is entropy refined by $\Delta_I$, written $\Delta_S \preceq \Delta_I$, just when there is some super $\pmb{\Delta}$ such that

$$\Delta_S = \pmb{\mu}.\pmb{\Delta} \quad \wedge \quad (\mathcal{B}\pmb{\mu}).\pmb{\Delta} = \Delta_I \ . \tag{49}$$

□

## B.2 Notation for integration over measures

Because the coming calculations are intricate in places, we use a slightly non-standard notation for integration over measures in order to make the manipulation of bound variables etc. absolutely explicit.[24] Fix a measure space $(\pmb{A}, \mathcal{A}, \alpha)$, with $\pmb{A}$ the underlying space and $\mathcal{A}$ a sigma-algebra on it, and $\alpha$ a measure. We consider the expression "$exp\, \mathrm{d}a$" simply to be an alternative notation for the lambda expression $(\lambda a \cdot exp)$, i.e. with $\mathrm{d}a$ binding free occurrences of $a$ within $exp$ to make a function over $\pmb{A}$. (Note this convention accords perfectly with the notation for Riemann integration, in particular that the "$\mathrm{d}a$" binds occurrences of $a$ in the body.) Then $\int_\alpha^A exp\, \mathrm{d}a$ means

> Consider the expression $exp$ to be a function of its free variable $a$, and integrate that function over the measure $\alpha$, but restricted to the measurable set $A$ in $\mathcal{A}$.

When $f$ say actually *is* some function over $\pmb{A}$ (rather than an expression containing $a$), then we write just $\int_\alpha^A f$ with no indication "d·" of bound variable, and it is of course equivalent to $\int_\alpha^A f.a\, \mathrm{d}a$ where $exp$ is now the function application $f.a$. Finally, when there is no restricting set $A$ we can leave it off. Thus the simplest "normally notated" integration $\int f \mathrm{d}\mu$ we would write instead as $\int_\mu f$.[25]

---

[24]This is not done lightly: variant notation always imposes a barrier between writer and reader. As justification in this case, we quote "Sometimes the integral of a function $h$ with respect to a measure $\mu$, usually written as $\int h\mathrm{d}\mu$ or $\int h(x)\mathrm{d}\mu(x)$, will be written as $\int h(x)\mu(\mathrm{d}x)$. This can make clearer what the variable of integration is... [14, p347]." In the last case, what seems to be meant is that $\int h(x)\mu(\mathrm{d}x, y)$ would be an integration of function $h$ of $x$ over a measure $\mu$ on that $x$ with $\mu$ depending on some $y$ — which we would write in our notation as $\int_{\mu(y)} h(x)\mathrm{d}x$.

As an example of how this can get out of hand, consider the expressions

$$\int_{X_1} \mu_{x_1} \mathrm{d}(\pi_1)*(\mu)(x_1)$$
and $\int_{X_1} \left( \int_{X_2} f(x_1, x_2)\mu(\mathrm{d}x_2|x_1) \right) \mu(\pi_1^{-1}(\mathrm{d}x_1))\ ,$

taken from the product-space section of Wiki entry on the Disintegration Theorem [1, .../Disintegration_theorem]. It's reminiscent of "Von Neumann's onion"

$$(\psi((((a)))))^2 \quad = \quad \phi((((a))))\ ,$$

so called because "it has to be peeled before it can be digested" [17, *The legend of John von Neumann*].

[25]This avoids all the contortions one reads like $\mathrm{d}\mu(x)$ and $\mu(\mathrm{d}x)$ as ad-hoc variations on the "normal" use, varying from text to text depending on the complexity of their calculations.

### B.3 Convex functions and Jensen's inequality

We will use a *convex* function y on $\mathbb{D}D$, i.e. one with the property that for $0 \leq p \leq 1$ and $\delta_{\{1,2\}}: \mathbb{D}\mathcal{D}$ we have $y.(\delta_1 {}_p\oplus \delta_2) \geq y.\delta_1 {}_p\oplus y.\delta_2$. It is *strictly* convex if the inequality is strict whenever $\delta_1 \neq \delta_2$ and $p \neq 0, 1$. We define strictly convex $y.\delta := \sum_d (\delta.x)^2$, motivated by the "colour" construction for inhomogeneous Markov chains [45].

From Jensen's inequality [31, Thm.2] we have for convex y and hyper $\Delta$ that $\int_\Delta y \geq y.(\mathsf{avg}.\Delta)$. Defining $\mathsf{Y}.\Delta := \int_\Delta y$, we write this $\mathsf{Y}.\Delta \geq y.(\boldsymbol{\mu}.\Delta)$.

### B.4 Entropy refinement does not increase Y

We consider for $\Delta_{\{S,I\}}$ in $\mathbb{M}\mathbb{D}D$ the entropy refinement $\Delta_S \preceq \Delta_I$, beginning with the first of the two criteria from Def. 5.1 that there be a super $\boldsymbol{\Delta}$ in $\mathbb{M}^2\mathbb{D}D$ with $\Delta_S = \mathsf{avg}.\boldsymbol{\Delta}$. We calculate

$$\begin{aligned}
& \mathsf{Y}.\Delta_S \\
=\ & \int_{\Delta_S} y \\
=\ & \int_{\mathsf{avg}.\boldsymbol{\Delta}} y \\
=\ & \int_{\boldsymbol{\Delta}} (\int_\Delta y)\, d\boldsymbol{\Delta} && \text{"[15, Thm.1(d)]"} \\
=\ & \int_{\boldsymbol{\Delta}} \mathsf{Y}.\Delta\, d\boldsymbol{\Delta} \\
=\ & \int_{\boldsymbol{\Delta}} \mathsf{Y}.
\end{aligned}$$

From the second criterion of refinement, that $\mathsf{map}.\mathsf{avg}.\boldsymbol{\Delta} = \Delta_I$, we calculate

$$\begin{aligned}
& \mathsf{Y}.\Delta_I \\
=\ & \mathsf{Y}.(\mathsf{map}.\mathsf{avg}.\boldsymbol{\Delta}) \\
=\ & \int_{\mathsf{map}.\mathsf{avg}.\boldsymbol{\Delta}} y \\
=\ & \int_{\boldsymbol{\Delta}} y \circ \mathsf{avg} && \text{"[15, Thm.1(a)]"} \\
=\ & \int_{\boldsymbol{\Delta}} y.(\mathsf{avg}.\Delta)\, d\Delta \\
\leq\ & \int_{\boldsymbol{\Delta}} \mathsf{Y}.\Delta\, d\Delta && \text{"y is convex"} \\
=\ & \int_{\boldsymbol{\Delta}} \mathsf{Y}.
\end{aligned}$$

Putting the two calculations together gives us $\mathsf{Y}.\Delta_S \geq \mathsf{Y}.\Delta_I$ immediately.

### B.5 Conditions for strict decrease of Y

The step $\int_{\boldsymbol{\Delta}} y.(\mathsf{avg}.\Delta)\, d\Delta \leq \int_{\boldsymbol{\Delta}} \mathsf{Y}.\Delta\, d\Delta$ depends only on the underlying inequality that $y.(\mathsf{avg}.\Delta) \leq \mathsf{Y}.\Delta$ for all $\Delta$, and so we will have equality there just when the set of $\Delta$'s such that $y.(\mathsf{avg}.\Delta) = \mathsf{Y}.\Delta$ has measure one in $\boldsymbol{\Delta}$. But $\Delta$ satisfies that equality precisely when it is of the form $\mathsf{pnt}.\delta$, that is $\{\!|\delta|\!\}$ for some $\delta$ in $\mathbb{D}\mathcal{X}$, i.e. is a point-hyper centred on that $\delta$ [31, Thm.5]. Thus we have equality iff the set $P := \{\!|\delta: \mathbb{D}\mathcal{X} \bullet \mathsf{pnt}.\delta|\!\}$ of point hypers in $\mathbb{M}\mathbb{D}\mathcal{X}$ has measure one in the super $\boldsymbol{\Delta}$.[26]

---

[26] $P$ is closed because it contains its limit points in the Kantorovich metric, hence is measurable.

But if $\mathbf{\Delta}.P{=}1$ then $\mathsf{avg}.\mathbf{\Delta} = \mathsf{map.avg}.\mathbf{\Delta}$,[27] whence $\Delta_S{=}\Delta_I$. That gives us that $\Delta_S{\preceq}\Delta_I$ implies $Y.\Delta_S{>}Y.\Delta_I$ unless $\Delta_S{=}\Delta_I$, whence antisymmetry of entropy refinement follows trivially.

## B.6 Antisymmetry of secure refinement

This now follows easily from the above because $\Delta_S{\sqsubseteq}\Delta_I{\sqsubseteq}\Delta_S$ implies $\sum\Delta_S{\leq}\sum\Delta_I{\leq}\sum\Delta_S$, whence $\sum\Delta_S{=}\sum\Delta_I$ and so $\Delta_S{\preceq}\Delta_I{\preceq}\Delta_S$, thus finally $\Delta_S{=}\Delta_I$ from §B.5.

# C Transitivity of secure refinement

We prove transitivity of refinement for three domains of increasing sophistication, taking advantage of the similarities between them to make the structure of the proof clearer. Although direct matrix-based proofs are possible in the discrete case, our aim is to use monadic-style arguments that are easily generalised to measures.

## C.1 A useful conjecture

If we had the following property in our monad, transitivity of entropy refinement would be straightforward:

**Conjecture C.1** Suppose we have a hyper $\Delta$ in $\mathbb{D}^2\mathcal{X}$ and two supers $\mathbf{\Delta}_{\{1,2\}}$ in $\mathbb{D}^3\mathcal{X}$ with the property that $\mathsf{map.avg}.\mathbf{\Delta}_1{=}\Delta$ and $\Delta{=}\mathsf{avg}.\mathbf{\Delta}_2$. Then there is a "mega" $\nabla$ in $\mathbb{D}^4\mathcal{X}$ such that $\mathbf{\Delta}_1{=}\mathsf{avg}.\nabla$ and $\mathsf{map}^2.\mathsf{avg}.\nabla{=}\mathbf{\Delta}_2$. □

Fig. 4 further below gives a diagram of this relationship (but in more mathematical notation). With Conj. C.1 the proof of transitivity of ($\preceq$) would be

**Lemma C.1** *Refinement is transitive* If $\Delta_1{\preceq}\Delta_2$ and $\Delta_2{\preceq}\Delta_3$ for hypers $\Delta_{\{1,2,3\}}$ in $\mathbb{D}^2\mathcal{X}$, then $\Delta_1{\preceq}\Delta_3$.

---

[27]If $\mathbf{\Delta}.P{=}1$ then $\mathbf{\Delta}{=}\mathsf{map.pnt}.\Delta$ for some $\Delta$, whence

$$\begin{aligned}\mathsf{avg}.\mathbf{\Delta} &= \mathsf{avg}.(\mathsf{map.pnt}.\Delta) = (\mathsf{avg} \circ \mathsf{map.pnt}).\Delta = \Delta \\ &= \mathsf{map}.(\mathsf{avg} \circ \mathsf{pnt}).\Delta = \mathsf{map.avg}.(\mathsf{map.pnt}.\Delta) = \mathsf{map.avg}.\mathbf{\Delta}\ .\end{aligned}$$

Alternatively, for measurable set $Q{\subseteq}P$ of hypers we calculate directly

$$\begin{aligned}&\mathbf{\Delta}.Q \\ =\ &\mathbf{\Delta}.\{\Delta{:}\,P \bullet (\mathsf{pnt} \circ \mathsf{avg}).\Delta \in Q\} &&\text{``}Q{\subseteq}P;\ \mathsf{pnt\circ avg}\text{ is identity on }P\text{''} \\ =\ &\mathbf{\Delta}.\{\Delta{:}\,\mathbb{M}\mathcal{X} \bullet (\mathsf{pnt} \circ \mathsf{avg}).\Delta \in Q\} &&\text{``}\mathbf{\Delta}.P{=}1\text{''} \\ =\ &\mathsf{map}.(\mathsf{pnt} \circ \mathsf{avg}).\mathbf{\Delta}.Q\ ,\end{aligned}$$

whence because $\mathbf{\Delta}.P{=}1$ we can ignore the assumption $Q{\subseteq}P$ to conclude that $\mathbf{\Delta} = \mathsf{map}.(\mathsf{pnt{\circ}avg}).\mathbf{\Delta}$, i.e. that $\mathbf{\Delta} = \mathsf{map.pnt}.\Delta_I$. Then $\Delta_S = \mathsf{avg}.\mathbf{\Delta} = \Delta_I$.

*Proof:* From Def. 5.1 and Conj. C.1 we have $\mathbf{\Delta}_{\{1,2\}}$ in $\mathbb{D}^3\mathcal{X}$ and $\mathbf{\nabla}$ in $\mathbb{D}^4\mathcal{X}$ with

$$\Delta_1 = \mathsf{avg}.\mathbf{\Delta}_1 \quad \wedge \quad \mathsf{map.avg}.\mathbf{\Delta}_1 = \Delta_2$$
$$\Delta_2 = \mathsf{avg}.\mathbf{\Delta}_2 \quad \wedge \quad \mathsf{map.avg}.\mathbf{\Delta}_2 = \Delta_3$$
$$\mathbf{\Delta}_1 = \mathsf{avg}.\mathbf{\nabla} \quad \wedge \quad \mathsf{map}^2.\mathsf{avg}.\mathbf{\nabla} = \mathbf{\Delta}_2 \ .$$

Monad laws then give $\Delta_1 = \mathsf{avg}.(\mathsf{map.avg}.\mathbf{\nabla})$ and $\mathsf{map.avg}.(\mathsf{map.avg}.\mathbf{\nabla}) = \Delta_3$, so that $\Delta_1 \preceq \Delta_3$ is established with witness $\mathsf{map.avg}.\mathbf{\nabla}$. □

That leaves of course the proof of Conj. C.1, so there is still some work to do: we explore the conjecture in three stages.

## C.2 Proof of Conj. C.1 for a qualitative model

Here we attempt the proof of our conjecture in the *qualitative* model of non-interference and refinement [35, 36, 38], i.e. for sets instead of distributions. (Refinement is already known to be transitive for that model; we are redoing it in a different way in order to bolster our intuition for its generalisation.)

For succinctness, we will from here on use more conventional monad-notation: the functor (map) is $\mathbb{M}$ and the multiply transformation (avg) is $\boldsymbol{\mu}$.

The first move it to make the conjecture slightly more general, hence in fact simpler: we assume we have $\mathbb{M}f.X = B = \boldsymbol{\mu}.Y$ for some $B, X, Y$ and $f$ all of the right types. We will find $Z$ such that $\boldsymbol{\mu}.Z{=}X$ and $\mathbb{M}^2 f.Z{=}Y$. Using a general $f$, instead of $\boldsymbol{\mu}$ as in Conj. C.1 specifically, means we can think one level lower than before: we now have that $B, X$ are just sets, and $Y$ –and $Z$ (effectively the $\mathbf{\nabla}$ of Conj. C.1, as we will see)– are simply sets of sets.

With this setup, the construction of $Z$ is very straightforward: it's the set of sets $\{y{:}Y \bullet \{x{:}X \mid f.x{\in}y\}\}$. We calculate first

$$\begin{aligned}
& \boldsymbol{\mu}.Z \\
={} & \{z{:}Z; x'{:}z \bullet x'\} \\
={} & \{z{:}\{y{:}Y \bullet \{x{:}X \mid f.x{\in}y\}\}; x'{:}z \bullet x'\} & \text{``defn } Z\text{''} \\
={} & \{y{:}Y; x'{:}\{x{:}X \mid f.x{\in}y\} \bullet x'\} \\
={} & \{y{:}Y; x{:}X \mid f.x{\in}y \bullet x\} \\
={} & \{x{:}X \mid (\ \exists y{:}Y \mid f.x{\in}y\ )\} \\
={} & X \ . & \text{``}\mathbb{M}f.X = \boldsymbol{\mu}.Y\text{''}
\end{aligned}$$

The other calculation is

$$\begin{aligned}
& \mathbb{M}^2 f.Z \\
={} & \{y{:}Y \bullet \{x{:}X \mid f.x{\in}y \bullet f.x\}\} & \text{``defn } Z\text{''} \\
={} & \{y{:}Y \bullet \{x{:}X; b \mid b{=}f.x \wedge b{\in}y \bullet b\}\} & \text{``one-point rule''} \\
={} & \{y{:}Y \bullet \{b{:}y \mid (\ \exists x{:}X \mid b{=}f.x\ )\}\} \\
={} & \{y{:}Y \bullet y\} & \text{``}\mathbb{M}f.X = \boldsymbol{\mu}.Y\text{''} \\
={} & Y \ .
\end{aligned}$$

That indeed proves Conj. C.1 for the qualitative model if we instantiate $f$ to $\boldsymbol{\mu}$. Based on it, we now turn to the discrete *quantitative* model, as in this report.

## C.3 Proof of Conj. C.1 for a quantitative model

Again we generalise, so that $\mathbb{M}f.X = B = \boldsymbol{\mu}.Y$ with $B, X$ distributions and $Y, Z$ hypers.

This time the construction of $Z := \{\!\{y{:}Y \bullet \{\!\{x{:}X \mid y.(f.x)/B.(f.x)\}\!\}\}\!\}$, where we are able to exploit the (deliberate) similarity of the notations in §8.1 with the ordinary, established comprehension- and enumeration notation of set theory that we used in §C.2 just above. We calculate first

$\quad \boldsymbol{\mu}.Z$
$= \ \{\!\{z{:}Z; x'{:}z \bullet x'\}\!\}$
$= \ \{\!\{z{:}\{\!\{y{:}Y \bullet \{\!\{x{:}X \mid y.(f.x)/B.(f.x)\}\!\}\}\!\}; x'{:}z \bullet x'\}\!\}$     "defn $Z$"
$= \ \{\!\{y{:}Y; x'{:}\{\!\{x{:}X \mid y.(f.x)/B.(f.x)\}\!\} \bullet x'\}\!\}$
$= \ \{\!\{y{:}Y; x'{:}(\odot x{:}X \bullet y.(f.x)/B.(f.x) \times \{\!\{x\}\!\}) \bullet x'\}\!\}$     "see below"
$= \ (\odot y{:}Y; x{:}X \bullet y.(f.x)/B.(f.x) \times \{\!\{x\}\!\})$
$= \ (\odot x{:}X \bullet \{\!\{x\}\!\} \times (\odot y{:}Y \bullet y.(f.x))/B.(f.x))$
$= \ (\odot x{:}X \bullet \{\!\{x\}\!\} \times B.(f.x)/B.(f.x))$     "$B = \boldsymbol{\mu}.Y$"
$= \ (\odot x{:}X \bullet \{\!\{x\}\!\})$
$= \ X \ .$

This calculation is not ideal: it tries to follow the calculation in §C.2, but it needs some extra steps. For the "see below" we calculate

$\quad (\odot x{:}X \bullet y.(f.x)/B.(f.x))$
$= \ (\odot b{:}B \bullet y.b/B.b)$     "$B = \mathbb{M}f.X$"
$= \ (\ \sum b{:}\lceil B \rceil \bullet y.b\ )$
$= \ 1\ ,$     "$y$ is total"

so that the denominator of the conditional comprehension can be removed. We use this same identity below for the second calculation, reasoning

$\quad \mathbb{M}^2 f.Z$
$= \ \{\!\{y{:}Y \bullet \{\!\{x{:}X \mid y.(f.x)/B.(f.x) \bullet f.x\}\!\}\}\!\}$     "defn $Z$"
$= \ \{\!\{y{:}Y \bullet (\odot x{:}X \bullet y.(f.x)/B.(f.x) \times \{\!\{f.x\}\!\})\}\!\}$     "see above"
$= \ \{\!\{y{:}Y \bullet (\odot b{:}B \bullet y.b/B.b \times \{\!\{b\}\!\})\}\!\}$     "$B = \mathbb{M}f.X$"
$= \ \{\!\{y{:}Y \bullet y\}\!\}$
$= \ Y \ .$

That proves Conj. C.1 for the discrete-distribution case, sufficient in fact for this report. We now turn to proper measures.

## C.4 Refinement for proper measures

### C.4.1 Product measures and conditionals

We begin with three measurable spaces $(\boldsymbol{A}, \mathcal{A})$ and $(\boldsymbol{B}, \mathcal{B})$ with $(\boldsymbol{G}, \mathcal{G})$ being their product. The *Product Measure Theorem* [4, p97] assumes a measure $\alpha$ and a function $f{:}\boldsymbol{A}{\to}\mathbb{M}\boldsymbol{B}$, that is equivalently of type $\boldsymbol{A}{\to}\boldsymbol{\mathcal{B}}{\to}\mathbb{R}$, with the further property that for any fixed $B{\in}\mathcal{B}$ the function $f.(\cdot).B$ is measurable on $\mathcal{A}$ so that integrations $\int_\alpha (f.a.B)\mathrm{d}a$ are meaningful. (There are also various assumptions about the measures' being finite, which we satisfy because we're using probability measures, thus bounded by 1.)

Under those assumptions there is a unique product measure $\gamma$ on $\mathcal{G}{:=}\mathcal{A}{\times}\mathcal{B}$ satisfying [28]

$$\gamma(A{\times}B) \quad = \quad \int_\alpha^A f.a.B \, \mathrm{d}a \quad \text{for all measurable } A,B \text{ in } \mathcal{A},\mathcal{B}, \qquad (50)$$

given by $\gamma.G{:=}\int_\alpha f.a.G_a \, \mathrm{d}a$ where $G_a$ is the "section" $\{b{:}\boldsymbol{B} \mid (a,b){\in}G\}$ of $G$ at $a$.

The function $f$ is like a conditional probability for (the constructed) product measure $\gamma$, giving for each (second-coordinate measurable set) $B$ a probability distribution conditioned on (first-coordinate point) $a$. [29]

The *Disintegration Theorem* [1, `.../Disintegration_theorem`] goes in the other direction. (It's a specialisation of techniques for *Regular Conditional Probabilities* [4, §6.6], [14, Prop. 10.2.8].) Here –in essence– we begin with a $\gamma$ as above and try to find a suitable $f$ that satisfies (50). To work, it requires that $(\boldsymbol{A}, \mathcal{A})$ and $(\boldsymbol{B}, \mathcal{B})$ be *Radon* spaces.

The theorem says that for any measure $\gamma$ on the product $\mathcal{G} = \mathcal{A}{\times}\mathcal{B}$ there is an $f{:}\boldsymbol{A}{\to}\mathbb{M}\underline{B}$ depending on $\gamma$ such that (50) holds with $\alpha{:=}\overleftarrow{\gamma}$ being the marginal measure of $\gamma$ on its first coordinate, that is given by $\alpha.A = \gamma(A{\times}\boldsymbol{B})$. Furthermore, this $f$ is "almost uniquely determined" in the sense that any other $f'$ satisfying (50) agrees with $f$ except possibly on a subset of $\boldsymbol{A}$ with $\alpha$-measure zero: we will write that as $f \approx_\alpha f'$, meaningful only when $f, f'$ are functions on $\boldsymbol{A}$.

For notational convenience, given some $\gamma$ we'll write $\overleftarrow{\gamma}$ for such an $f$ (a member of the equivalence class), being careful in that case to use only operations on $\overleftarrow{\gamma}$ for which the class is a congruence; similarly we write $\overrightarrow{\gamma}$ for such a function on $B$. Thus we have for any $\gamma$ and $\alpha{=}\overleftarrow{\gamma}$ and $\beta{=}\overrightarrow{\gamma}$ the existence of $\overleftarrow{\gamma}$ and $\overrightarrow{\gamma}$ such that for all measurable sets $A, B$ in $\mathcal{A}, \mathcal{B}$ we have

$$\gamma.(A{\times}B) \quad = \quad \int_\alpha^A \overleftarrow{\gamma}.a.B \quad = \quad \int_\beta^B \overrightarrow{\gamma}.b.A \ . \qquad (51)$$

(We note that the integrations are indeed congruential operations.) In the $\overrightarrow{\gamma}$ case the order of arguments is $b, A$ (rather than $a, B$), so that the point always comes first and then the measurable set. [30]

---

[28] Note that this product is not Cartesian: it's a product of sigma-algebras, thus the smallest sigma-algebra *containing* the Cartesian product.

[29] By analogy, in the discrete case we'd have $\alpha$ as a column vector on the left and, for each row-index $a$ the function $f.a$ (that is $f(a, \cdot)$) would be a normalised row for that index. The induced "denormalised" row $\alpha.a \times f.a$ then gives the actual row for that index $a$, and all those rows piled up together give the matrix for the product distribution $\gamma$.

[30] For discrete measures this is saying that if we have a joint-distribution matrix $\gamma$ over $\boldsymbol{G} = \boldsymbol{A}{\times}\boldsymbol{B}$ then it can be presented as an $a$-indexed collection of normalised rows $\overleftarrow{\gamma}.a$ and also as a $b$-indexed collection of normalised columns $\overrightarrow{\gamma}.b$, i.e. either way as we prefer.

As an example of the above, we give two lemmas that we'll need later. Recall that (bold) $\boldsymbol{\mu}$ is multiplication from the probabilistic monad (i.e. it is not some measure $\mu$): it "averages" a measure of measures to give a single measure again.

The first lemma would say in discrete terms that if you have a joint distribution matrix $\varepsilon$ over $\boldsymbol{E}=\boldsymbol{D}\times\boldsymbol{B}$ with marginals $\delta,\beta$, and you "relabel" the right-marginal $\beta$ by naming its columns by the columns' *values* themselves, giving a relabelled distribution $\zeta$, then in fact $\zeta$ averages to $\delta$. This actually is how you can convert a hyper presented via an index-set ($\boldsymbol{B}$) into a "real" hyper given directly as a measure of measures, i.e. with no index-set having to be defined separately.

**Lemma C.2** *Average of conditionals*   Let $\varepsilon$ be a joint measure over $\mathcal{E} = \mathcal{D}\times\mathcal{B}$ with marginals $\delta=\overleftarrow{\varepsilon}$ and $\beta=\overrightarrow{\varepsilon}$. Note that $\overrightarrow{\varepsilon}$ is of type $\underline{B}\to\mathbb{M}\underline{D}$ so that $\mathbb{M}\overrightarrow{\varepsilon}$ is of type $\mathbb{M}\underline{B}\to\mathbb{M}^2\underline{D}$, and construct the measure $\zeta:=\mathbb{M}\overrightarrow{\varepsilon}.\beta$ in $\underline{Z}:=\mathbb{M}^2\underline{D}$.

Then we have $\delta=\boldsymbol{\mu}.\zeta$ .

*Proof:*   Calculate for any measurable $D$ in $\mathcal{D}$ that

$$\begin{aligned}
&\boldsymbol{\mu}.\zeta.D \\
=\quad & \int_\zeta z.D\,dz & \text{"defn $\boldsymbol{\mu}$ (and assuming evaluation function $(.D)$ is measurable)"} \\
=\quad & \int_{\mathbb{M}\overrightarrow{\varepsilon}.\beta} z.D\,dz & \text{"defn $\zeta$"} \\
=\quad & \int_\beta (\overrightarrow{\varepsilon}.b).D\,db & \text{"chain rule for $\mathbb{M}\overrightarrow{\varepsilon}$"} \\
=\quad & \varepsilon.(D\times\boldsymbol{B}) & \text{"$\beta=\overrightarrow{\varepsilon}$, and (51) for $\overrightarrow{\varepsilon}$"} \\
=\quad & \overleftarrow{\varepsilon}.D & \text{"defn marginal"} \\
=\quad & \delta.D\ , & \text{"defn $\delta$"}
\end{aligned}$$

which suffices, since $D$ was arbitrary.   $\square$

The second lemma concerns "mapping" of a joint distribution's conditionals. It would say in discrete terms that if you have a joint distribution matrix $\gamma$ over $\boldsymbol{G}=\boldsymbol{A}\times\boldsymbol{B}$ with marginals $\alpha,\beta$, but in fact the left-marginal $\alpha$ is given as a "pushforward measure" $\mathbb{M}f.\delta$ via $f\colon \boldsymbol{D}\to\boldsymbol{A}$ from some other $\delta$, then you can make a new joint distribution $\varepsilon$ over $\boldsymbol{D}\times\boldsymbol{B}$ that relates $\delta,\beta$ directly, i.e. so that $\delta=\overleftarrow{\varepsilon}$ and $\beta=\overrightarrow{\varepsilon}$ and –moreover– the columns of $\varepsilon$, distributions themselves over $\delta$, map (push-forward) via $f$ to the corresponding columns of the original $\gamma$.

**Lemma C.3** *Push-forward of conditionals*   Let $\gamma$ be a joint measure over the sigma-algebra $\mathcal{G} = \mathcal{A}\times\mathcal{B}$ with marginals $\alpha=\overleftarrow{\gamma}$ and $\beta=\overrightarrow{\gamma}$, and let there be a further measure $\delta$ over $\mathcal{D}$ and function $f\colon \boldsymbol{D}\to\boldsymbol{A}$ such that $\alpha=\mathbb{M}f.\delta$.

Then there is a joint measure $\varepsilon$ over $\underline{E}=\underline{D}\times\underline{B}$ with marginals $\delta,\beta$ such that we have $\overrightarrow{\gamma}\approx_\beta \mathbb{M}f\circ\overrightarrow{\varepsilon}$.[31]

*Proof:*   Define $\overleftarrow{\varepsilon}:=\overleftarrow{\gamma}\circ f$, and observe trivially for the joint distribution $\varepsilon$ in $\mathbb{M}(\underline{E}\times\underline{B})$ induced via the Product Measure Theorem that indeed $\delta=\overleftarrow{\varepsilon}$ and $\beta=\overrightarrow{\varepsilon}$.

---

[31] Recall that $(\approx_\beta)$ means "equal except on a set of $\beta$-measure zero, and that $(\circ)$ is functional composition.

We now use the *almost-unique* property of $\vec{\gamma}$, calculating for arbitrary $A\colon\mathcal{A}$ and $B\colon\mathcal{B}$ that

$$
\begin{aligned}
& \int_\beta^B (\mathbb{M}f\circ\vec{\varepsilon}).b.A\,\mathrm{d}b \\
={} & \int_\beta^B \mathbb{M}f.(\vec{\varepsilon}.b).A\,\mathrm{d}b & \text{"composition"} \\
={} & \int_\beta^B \vec{\varepsilon}.b.(f^{-1}.A)\,\mathrm{d}b & \text{"defn } \mathbb{M}f\text{"} \\
={} & \varepsilon.((f^{-1}.A)\times B) & \text{"defn } \vec{\varepsilon}\text{"} \\
={} & \int_\delta^{f^{-1}.A} \cev{\varepsilon}.d.B\,\mathrm{d}d & \text{"property of } \cev{\varepsilon}\text{ wrt } \varepsilon\text{"} \\
={} & \int_\delta^{f^{-1}.A} \cev{\gamma}.(f.d).B\,\mathrm{d}d & \text{"defn } \cev{\varepsilon}\text{"} \\
={} & \int_\alpha^A \cev{\gamma}.a.B\,\mathrm{d}a & \text{"chain rule, since } \alpha=\mathbb{M}f.\delta\text{"} \\
={} & \gamma.(A\times B)\,, & \text{"property of } \cev{\gamma}\text{ wrt } \gamma\text{"}
\end{aligned}
$$

which is the defining property up to $\approx_\beta$ of $\vec{\gamma}$. □

### C.4.2 Proof of Conj. C.1

With the above preparation, we can now prove Conj. C.1 for general measures. We restate it using the notational conventions of this section:

*Conjecture C.1′* Suppose we have a measure $\alpha$ and two other measures $\delta,\beta$ with the properties that $\mathbb{M}f.\delta=\alpha$ and $\alpha=\boldsymbol{\mu}.\beta$, where $f\colon\boldsymbol{D}\to\boldsymbol{A}$.

Then there is a measure $\zeta$ in $\mathbb{M}^2\underline{D}$ such that $\delta=\boldsymbol{\mu}.\zeta$ and $\beta=\mathbb{M}^2f.\zeta$. □

Fig. 4 gives the relevant commuting diagram. Note however that the arrows ($\mapsto$) are *applied* to the vertices — they are not functions *between* the vertices. Here is the proof:

1. Because $\alpha=\boldsymbol{\mu}.\beta$ we can construct a joint $\gamma$ over $\underline{G}=\underline{A}\times\underline{B}$ with marginals $\alpha,\beta$ such that $\vec{\gamma}=1$, the identity function on $\boldsymbol{B}$. This follows directly from Lem. C.2, since $\mathbb{M}1=1$. [32]

2. Note at this point –just for keeping things straight– that since $\alpha=\boldsymbol{\mu}.\beta$ in fact $\boldsymbol{B}=\mathbb{M}\underline{A}$, that is elements $b\colon\boldsymbol{B}$ can be applied to measurable subsets $A\colon\mathcal{A}$ of $\boldsymbol{A}$.

3. Now, as in Lem. C.3, given our assumption $\alpha=\mathbb{M}f.\delta$, construct joint $\varepsilon$ over $\mathcal{E}=\mathcal{D}\times\mathcal{B}$ with marginals $\delta,\beta$ and such that $\vec{\gamma}\approx_\beta \mathbb{M}f\circ\vec{\varepsilon}$.

4. And now, as in Lem. C.2, construct measure $\zeta:=\mathbb{M}\vec{\varepsilon}.\beta$. (For the types here, remember that $\beta\in\mathbb{M}\underline{B}=\mathbb{M}^2\underline{A}$, and $\vec{\varepsilon}\colon\boldsymbol{B}\to\mathbb{M}\underline{D}$, and that's why $\zeta=\mathbb{M}\vec{\varepsilon}.\beta$ is of type $\mathbb{M}^2\underline{D}$.)

5. Observe that we then have $\delta=\boldsymbol{\mu}.\zeta$ directly from Lem. C.2.

---

[32] ...or it can be calculated directly: define $\vec{\gamma}.b.A:=b.A$ and then verify that indeed we have $\cev{\gamma}=\boldsymbol{\mu}.\beta$.

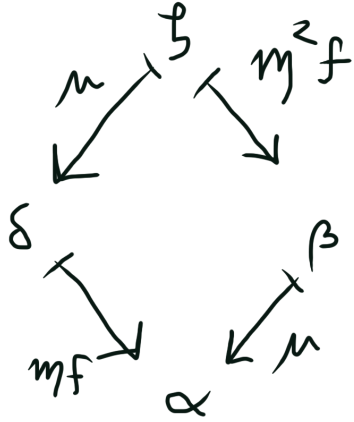

Conjecture $C.1'$ establishes the existence of the upper half, a $\zeta$ such that $\delta=\boldsymbol{\mu}.\zeta$ and $\mathbb{M}^2 f.\zeta=\beta$, given the lower half $\mathbb{M}f.\delta = \alpha = \boldsymbol{\mu}.\beta$.

Figure 4: Conjecture $C.1'$

6.  Finally, calculate

$$\begin{aligned}
&\mathbb{M}^2 f.\zeta \\
=\ &\mathbb{M}^2 f.(\mathbb{M}\vec{\varepsilon}.\beta) &&\text{``defn }\zeta\text{''} \\
=\ &\mathbb{M}(\mathbb{M}f\circ\vec{\varepsilon}).\beta &&\text{``functor''} \\
=\ &\mathbb{M}\vec{\gamma}.\beta &&\text{``(3.) above gives }\mathbb{M}f\circ\vec{\varepsilon} \approx_\beta \vec{\gamma}\text{; see below''} \\
=\ &\beta\ . &&\text{``(1.) above: }\mathbb{M}\vec{\gamma}=\mathbb{M}\mathbf{1}=\mathbf{1}\text{''}
\end{aligned}$$

For the *see below* we note that both $\mathbb{M}f\circ\vec{\varepsilon}$ and $\vec{\gamma}$ are functions on $\boldsymbol{B}$ (in fact of type $\boldsymbol{B}\to\boldsymbol{B}$), and that in general if we have two functions $f_{\{1,2\}}:\boldsymbol{P}\to\boldsymbol{F}$ with $f_1 \approx_\pi f_2$ for some $\pi:\mathbb{M}\underline{P}$, then $\mathbb{M}f_1.\pi=\mathbb{M}f_2.\pi$.[33]

That establishes the transitivity of entropy refinement in the general case of proper measures.

## D  Refinement chains have suprema

As we showed in §6.1, the discrete hypers are not closed under suprema of chains: for the example we gave there, a measure was required. We show here that in

---

[33]Reason that for any $F\in\mathcal{F}$ we have

$$\begin{aligned}
&\mathbb{M}f_1.\pi.F \\
=\ &\pi.(f_1^{-1}.F) \\
=\ &\pi.(f_2^{-1}.F) &&\text{``}f_1 \approx_\pi f_2\text{''} \\
=\ &\mathbb{M}f_2.\pi.F\ ,
\end{aligned}$$

hence $\mathbb{M}f_1.\pi = \mathbb{M}f_2.\pi$ since $F$ was arbitrary.

fact measures are sufficient, in other words that sup-closure is achieved.

**Definition D.1** *Continuous relation*  Say that a relation $R: X \leftrightarrow Y$ between two complete metric spaces $X, Y$ is *continuous* if for every pair of convergent sequences $\{x_i\}_i$ and $\{y_i\}_i$ with $x_i(R)y_i$ for all $i$ we have also

$$\lim_i x_i \ (R) \ \lim_i y_i \ .$$

□

**Lemma D.1** *Continuous functions as continuous relations*  If $f: X \rightarrow Y$ is continuous, then both $f$ and $f^{-1}$ are continuous relations.
*Proof:*  Immediate from Def. D.1. □

**Lemma D.2** *Composition of continuous relations*  If $R: X \leftrightarrow Y$ and $S: Y \leftrightarrow Z$ are continuous relations between metric spaces $X, Y, Z$ with additionally $Y$ compact, then their composition $R \circ S \in X \leftrightarrow Z$ is continuous also.
*Proof:*  With the assumptions of Def. D.1 wrt $R \circ S$ there is a sequence $\{y_i\}_i$ in $Y$ such that $x_i(R)y_i \wedge y_i(S)z_i$ for all $i$. From compactness of $Y$ there is then a convergent subsequence $\{\hat{y}_j\}_j$ with limit $\hat{y}$ say; and by continuity of $R, S$ separately the corresponding subsequences $\{\hat{x}_j\}_j, \{\hat{z}_j\}_j$ in $X, Z$ with limits $\hat{x}, \hat{z}$ satisfy $\hat{x}(R)\hat{y} \wedge \hat{y}(S)\hat{z}$ so that in fact $\hat{x}(R \circ S)\hat{z}$. □

**Lemma D.3** *Entropy refinement is a partial order*
*Proof:*  Its reflexivity is immediate by taking $\mathbf{\Delta} = (\mathbb{M}\boldsymbol{\eta}).\Delta_S$ in Def. B.1(6); its antisymmetry was proved in §B; its transitivity was proved in §C. □

**Lemma D.4** *Continuity of entropy refinement*  The entropy refinement relation ($\preceq$) between hypers, as defined in Def. B.1(6) and thus in $\mathbb{M}\mathbb{D}\mathcal{X} \leftrightarrow \mathbb{M}\mathbb{D}\mathcal{X}$, is continuous in the sense of Def. D.1. [34]
*Proof:*  From Lem. A.1 we have that $\boldsymbol{\mu}: \mathbb{M}^2\mathbb{D}\mathcal{X} \rightarrow \mathbb{M}\mathbb{D}\mathcal{X}$ is nonexpansive. Since also $\boldsymbol{\mu}: \mathbb{M}\mathbb{D}\mathcal{X} \rightarrow \mathbb{D}\mathcal{X}$ is nonexpansive we have that $\mathbb{M}\boldsymbol{\mu}: \mathbb{M}^2\mathbb{D}\mathcal{X} \rightarrow \mathbb{M}\mathbb{D}\mathcal{X}$ is nonexpansive as well (same lemma).

Since the entropy-refinement relation is the composition of the inverted first with the second, we have its continuity by Lem. D.1 and Lem. D.2 given that $\mathbb{M}^2\mathbb{D}\mathcal{X}$ is compact (and 1-bounded) because (ultimately) $\mathcal{X}$ is compact (and 1-bounded). □

**Lemma D.5** *Continuity of termination refinement*  The termination refinement relation ($\leq$) between hypers as defined in Def. 7.2 is continuous in the sense of Def. D.1.
*Proof:*  Trivial. □

**Corollary D.1** *Continuity of secure refinement*  The secure refinement relation ($\sqsubseteq$) between hypers as defined in Def. 7.3 is continuous in the sense of Def. D.1.
*Proof:*  Immediate from Lemmas D.5, D.4 and D.2 since $\mathbb{M}\mathbb{D}\mathcal{X}$ is compact. □

---

[34]For consistency with the main report we use $\mathcal{X}$ rather than $\underline{X}$ here.

**Lemma D.6** *Refinement chains have suprema*    Let $\{\Delta_i\}_i$ be an $(\sqsubseteq)$-chain in $\mathbb{MD}\mathcal{X}$; then it has a $(\sqsubseteq)$-supremum in $\mathbb{MD}\mathcal{X}$.

*Proof:*    Since $\mathbb{MD}\mathcal{X}$ is compact there is an infinite subsequence $\{\hat{\Delta}_j\}_j$ converging to $\Delta$. For any $\Delta_i$ in our original sequence consider the tail of the infinite subsequence beginning beyond that point, that is $\{\hat{\Delta}_j\}_{j \geq j_0}$ for $j_0$ corresponding to a point at $i$ or beyond in the original sequence. Since we have $\Delta_i \sqsubseteq \hat{\Delta}_j$ for all $j \geq j_0$ we have by continuity of refinement Cor. D.1 arranged that also $\Delta_i \sqsubseteq \lim_{j \geq j_0} \hat{\Delta}_j = \Delta$.

Now suppose that $\Delta_i \sqsubseteq \Delta'$ for all $i$. That means in particular that $\hat{\Delta}_j \sqsubseteq \Delta'$ for all $j$ and thus again by continuity that $\Delta \sqsubseteq \Delta'$.

Hence $\Delta = \sqcup_i \Delta_i$ as required.    □

# E    Iteration is monotonic for secure refinement

It is trivial (and often assumed without comment) that iteration defined as a least fixed-point with respect to some partial order $(\sqsubseteq)$ is monotonic with respect to that same order in the iteration body: this is simply distribution of $(\sqsubseteq)$ through $(\sqsubseteq)$-suprema of chains. In our case however we use termination refinement $(\leq)$ for the chains producing least fixed-points, for reasons we explained in §6.2, yet we are interested in the monotonicity of secure refinement $(\sqsubseteq)$ wrt their suprema.

We show that in fact the $(\leq)$-chains can be treated as $(\sqsubseteq)$-chains, so that the above trivial argument then applies.

**Lemma E.1** *Termination-refinement chains converge*    Let $\{\Delta_i\}_i$ be a $(\leq)$-chain in $\mathbb{MD}\mathcal{X}$. Then it converges in the Kantorovich metric.

*Proof:*    Recall that we are using subprobability measures, and observe that for any $\Delta \leq \Delta'$ the Kantorovich distance between $\Delta$ and $\Delta'$ cannot exceed $\sum \Delta' - \sum \Delta$, and that difference of weights converges to zero along any $(\leq)$-chain.    □

**Lemma E.2** *Termination- and refinement-suprema $(\sqsubseteq)$-agree on termination-chains*    Let $\{\Delta_i\}_i$ be a $(\leq)$-chain in $\mathbb{MD}\mathcal{X}$. Then $\bigsqcup_i \Delta_i = \bigvee_i \Delta_i$.

*Proof:*    We prove $\bigsqcup_i \Delta_i \sqsubseteq \bigvee_i \Delta_i \sqsubseteq \bigsqcup_i \Delta_i$ and appeal to antisymmetry.

We know already that both suprema exist, and it is trivial from $(\leq) \subseteq (\sqsubseteq)$ that $\bigsqcup_i \Delta_i \sqsubseteq \bigvee_i \Delta_i$. The other direction $\bigvee_i \Delta_i \sqsubseteq \bigsqcup_i \Delta_i$ follows from continuity of $(\sqsubseteq)$ and the convergence of $\{\Delta_i\}_i$ as established in Lem. E.1.    □

**Lemma E.3** *Iteration is monotonic with respect to secure refinement*    Let $\{\Delta_i^{\{1,2\}}\}_i$ be two $(\leq)$-chains in $\mathbb{MD}\mathcal{X}$ with $(\leq)$-suprema $\Delta^{\{1,2\}}$ respectively, and suppose that $\Delta_i^1 \sqsubseteq \Delta_i^2$ for each $i$. Then we have also $\Delta^1 \sqsubseteq \Delta^2$.

*Proof:*    This is now trivial, since Lem. E.2 established that $\{\Delta_i^{\{1,2\}}\}_i$ considered as $(\sqsubseteq)$-chains have those same $\Delta^{\{1,2\}}$ as their $(\sqsubseteq)$-suprema.    □